\documentclass[preprint, superscriptaddress, showpacs,preprintnumbers,amsmath,amssymb,prb]{revtex4}
\usepackage{graphicx}

\begin{document}

\thispagestyle{empty}

\title{Investigation of the Casimir interaction
between two magnetic metals in comparison with
nonmagnetic test bodies}
\author{
A.~A.~Banishev\footnote{Present adress: Department of Chemistry,
University of Illinois, Urbana, Illinois 61801, USA}}
\affiliation{Department of Physics and
Astronomy, University of California, Riverside, California 92521,
USA}
\author{
G.~L.~Klimchitskaya}
\affiliation{Central Astronomical Observatory
at Pulkovo of the Russian Academy of Sciences,
St.Petersburg, 196140, Russia}
\author{
 V.~M.~Mostepanenko}
\affiliation{Central Astronomical Observatory
at Pulkovo of the Russian Academy of Sciences,
St.Petersburg, 196140, Russia}
\author{
U.~Mohideen
}
\affiliation{Department of Physics and
Astronomy, University of California, Riverside, California 92521,
USA}

\begin{abstract}
We present the complete results for the dynamic experiment on
measuring the gradient of the Casimir force between magnetic
(Ni-coated) surfaces of a plate and a sphere. Special
attention is paid to the description of some details of the
setup, its calibration, error analysis and background effects.
Computations are performed in the framework of the Lifshitz theory at
nonzero temperature with account of analytic corrections to
the proximity force approximation and of surface roughness
using both the Drude and the plasma model approaches.
The theory of magnetic interaction between a sphere and a plate
due to domain structure of their surfaces is developed for
both out-of-plane and in-plane magnetizations in the absence
and in the presence of spontaneous magnetization.
It is shown that in all cases the magnetic contribution to the
measured force gradients is much smaller than the total
experimental error. The comparison between experiment and
theory is done using the rigorous statistical method.
It is shown that the theoretical approach taking into account
dissipation of free electrons is excluded by the
data at  a 95\% confidence level. The approach
neglecting dissipation
is confirmed by the data at more than 90\% confidence level.
We prove that the results of experiments with Ni-Ni, Ni-Au
and Au-Au surfaces taken together cannot be reconciled with
the approach including free electrons dissipation by the introduction of any
unaccounted background force, either attractive or repulsive.

\end{abstract}

\maketitle

\section{Introduction}

The Casimir interaction is a version of the van der Waals
interaction\cite{1} when the separation distance between the
interacting bodies exceeds a few nanometers, and
relativistic effects make an important contribution.
The investigation of this phenomenon goes back to the seminal
 paper by Casimir\cite{2} which predicted that there is an attractive
force between two neutral parallel ideal metal plates in
vacuum. The Casimir force originates from the existence of
zero-point oscillations of the electromagnetic field and
thermal photons. Lifshitz\cite{3} developed the general
theory of the van der Waals and Casimir forces between plates
made of different materials based on the theory of electromagnetic
fluctuations. At the present time the Casimir effect is
investigated along with other quantum phenomena caused by fluctuating
electromagnetic field.\cite{4,5,6}
It has found increasing favor in numerous applications ranging
from condensed matter physics, atomic physics to elementary
particle physics, astrophysics and cosmology.\cite{7,8,9}
Much attention is given to measurements of the Casimir force
between two test bodies made of different materials.
Thanks to modern laboratory techniques using atomic force
microscopes (AFM) and micromachined oscillators it has been made
possible to measure the Casimir interaction to a high precision
at submicrometer separation distances (see reviews in
Refs.\cite{10,11,12}). In comparisons between experiment and
theory, some unexpected features in the interaction of quantum
fluctuations with matter have been found connected with the role
of conduction electrons which remain poorly understood up to the
present (see below in Secs.~VI and VII).

The original version of the Lifshitz theory\cite{3} describes
materials of the test bodies by means of a single quantity,
the frequency-dependent dielectric permittivity
$\varepsilon(\omega)$. In so doing the main physical observables,
such as the Casimir free energy and force, are most conveniently
expressed via $\varepsilon(i\xi_l)$ where the Matsubara frequencies
are $\xi_l=2\pi k_BTl/\hbar$, $k_B$ is the Boltzmann constant,
$T$ is the temperature, $l=0,\,1,\,2,\,\ldots\,$, and $\hbar$ is
the Planck constant.
The magnetic permeability of materials was assumed to be equal
to unity, $\mu(i\xi_l)=1$. This is justified for diamagnets whose
magnetic properties are characterized by the
relation\cite{13,14,15}
$|\mu(0)-1|\sim 10^{-5}$. For paramagnets consisting of
paramagnetic magnetizable microparticles with no intrinsic
magnetic moment (the Van Vleck polarization
paramagnetism\cite{16})
the magnetic properties are also negligibly small.
The same holds for paramagnets in the narrow sense which consist
of microparticles possessing an intrinsic (permanent) magnetic
moments whose interaction remains negligibly small even with the
decrease of temperature to absolute zero.\cite{13,14,15,16}
This allows one to conclude\cite{17} that in
most of cases the contribution of the magnetic properties to the
Casimir interaction is very small.
There is, however, the subset of paramagnets in the broad
sense called ferromagnets whose atoms possess strongly
interacting constituent magnetic moments below the temperature
of the magnetic phase transition (the Curie temperature
$T_C$). This results in large magnetic permeabilities at zero
Matsubara frequency, $\mu(0)\gg 1$, in the temperature region
$T<T_C$. Richmond and Ninham\cite{18} have generalized the
Lifshitz theory for the case of interacting bodies described by
the dielectric permittivity $\varepsilon(i\xi_l)$ and magnetic
permeability $\mu(i\xi_l)$ calculated at imaginary Matsubara
frequencies.

After generalization of the Lifshitz theory for the case of
magnetic plates, much theoretical work has been done.
Specifically, all main equations of the theory were
obtained\cite{19,20} for an arbitrary number of plane parallel
layers of magnetodielectrics possessing different
$\varepsilon(i\xi_l)$ and $\mu(i\xi_l)$. Furthermore, the
Lifshitz theory of van der Waals and Casimir interactions was
formulated for magnetodielectric bodies of arbitrary
shape.\cite{21} Many papers aimed to use magnetic properties
in order to realize the Casimir
repulsion.\cite{22,23,24,25,26,27,28,29}
It was understood,\cite{25,26} however, that
 $\mu(i\xi_l)$ decreases rapidly with $l$
in accordance with the Debye formula\cite{15}
\begin{equation}
\mu(i\xi_l)=1+\frac{\mu(0)-1}{1+\xi_l/\omega_m},
\label{Debye}\end{equation}
\noindent
where $\omega_m$ is the characteristic frequency which is much
less than $\xi_1\sim 10^{14}\,$Hz at room temperature.  For ferromagnetic metals
$\mu(i\xi)$ becomes equal to unity at $\xi> 10^5\,$Hz
(see, e.g., Ref.\cite{29a}).
From this it follows that
 the magnetic Casimir
interaction is determined by only the zero-frequency Matsubara
term  (i.e., the term with $l=0$ in the Lifshitz formula).
As a result, under some conditions the magnetic repulsion is
now expected only between two test bodies one of which is made
of ferromagnetic dielectric and another of a nonmagnetic
metal.\cite{24,25,26}
In parallel with the magnetic Casimir interaction between two
macroscopic bodies the case of polarizable microparticles
(atoms) with both electric and magnetic polarizabilities was
considered.\cite{29}
It was found that magnetic properies of both atoms and  material
of the wall influence the atom-wall interaction.\cite{19,30,31}

Recent Ref.\cite{32} marked the beginning of experimental research
of the magnetic Casimir interaction. In this experiment the
dynamic AFM operated in the frequency-shift mode was used to measure the
gradient of the Casimir force between an Au-coated sphere of
$64.1\,\mu$m radius oscillating in perpendicular direction to the
plate covered with the ferromagnetic metal Ni.
The dymanic AFM technique with a sharp tip has been used for
mapping surface topography for many years.\cite{33}
For measurements of the gradient of the Casimir force the dynamic
AFM was used in the phase-shift\cite{34,35} and in the
amplitude-shift\cite{36,37} modes.
When using the dynamic AFM in the frequency-shift mode, the
gradient of the Casimir force acting on the cantilever modifies
the resonant frequency and the corresponding frequency shift is
measured by means of a phase locked loop (PLL).
For AFM with a sharp tip this measurement mode was discussed
in detail in Ref.\cite{38}.
To measure the Casimir interaction by means of an AFM, it was
originally applied\cite{39} in the configuration of an Au-coated
sphere oscillating near an Au-coated plate. Previously dynamic
measurements of the Casimir interaction in the frequency-shift
mode were performed by means of a micromachined
oscillator.\cite{40,41,42,43,44,45,46,47}
Measurements of the Casimir interaction between an Au-coated
sphere and a Ni-coated plate\cite{32} demonstrated the impact
of magnetic properties of Ni, as predicted by the Lifshitz
theory with neglected relaxation properties of conduction electrons
(this theoretical approach was experimentally confirmed
previously by measurements with two Au test
bodies;\cite{39,42,43,44,45} see Secs.~VI and VII for a complete
discussion). However, with inclusion of the relaxation properties of free
charge carriers, the Lifshitz theory does not predict any impact
of magnetic properties on the Casimir interaction in the Au-Ni
configuration. Unfortunately, both theoretical predictions, by
coincidence, numerically almost overlap over the experimental
separation region. This does not allow to conclude that
Ref.\cite{32} alone
contains an independent confirmation for the impact of magnetic
properties on the Casimir interaction.

The convincing confirmation for the role of magnetic properties
in the Casimir effect was achieved\cite{48} by measuring the
gradient of the Casimir force between a Ni-coated sphere and
a Ni-coated plate by means  of dynamic AFM operated in the
frequency-shift mode. In this configuration the Lifshitz theory
predicts sufficiently different values of the gradient of the
Casimir force in cases when the relaxation properties of
conduction electrons are either included or neglected, and
in both cases the magnetic properties have a pronounced effect
on the result. Using the same setup, as in Refs.\cite{32} and
 \cite{39} for Au-Ni and Au-Au configurations, respectively,
it was shown that the magnetic properties of Ni affect the
measured gradient of the Casimir force.
The experimental results were found to be in excellent agreement
with the predictions of the Lifshitz theory with the
relaxation properties of free charge carriers neglected.
The theoretical predictions which take into account relaxation
properties of free electrons were experimentally excluded at
a high confidence level.\cite{48}
Remarkably, for Ni-Ni configuration the predictions of two
theoretical approaches change places, as compared to the case
of Au-Au test bodies.\cite{48} This leads to important
conclusions concerning the role of some possible background
effects (see Secs.~VI and  VII).

The present paper contains full description of the experiment
on measuring the gradient of the Casimir force between Ni-coated
surfaces of a sphere and a plate which was briefly described in
Ref.\cite{48}. After a necessary short discussion about the
measurement scheme (note that the setup is common for the
experiments of Refs.\cite{32,39} and  \cite{48}),
we present the measurement results including those which were
not published so far. The error analysis is elucidated in
more detail including the random, systematic and total
experimental errors. The Casimir interaction between two
Ni-coated surfaces used in this experiment is calculated
with the help of the Lifshitz theory within the two
theoretical approaches either neglecting or taking into
account the relaxation properties of free electrons.
In so doing the corrections due to surface roughness and
due to deviations from the proximity force approximation
(PFA) are taken into account.
Next, the detailed estimate of the magnetic interaction,
which might act in the experimental setup  due to the domain
structure of the films independent of the Casimir interaction,
is given. We demonstrate that the gradient of the
magnetic force is sufficiently small and cannot interfere in the
comparison between experiment and theory for both
cases of magnetization perpendicular or parallel to
the plane of the film.
Then the obtained experimental results are compared with the
results of numerical computations using the two theoretical
approaches. This is done with the help of a more rigorous
statistical method which was not used in
Refs.\cite{32,39} and  \cite{48}.
We arrive at the conclusion that the Lifshitz theory with
omitted relaxation properties of free electrons is
consistent with the measurement data whereas the same
theory with the inclusion of relaxation properties is excluded
by the data at a 95\% confidence level.
At the end of the paper we compare the experimental results of
this experiment with experiments of Refs.\cite{32} and  \cite{39}
involving at least one nonmagnetic (Au) surface.

The structure of the paper is as follows. In Sec.~II we briefly
present the measurement scheme using the dynamic AFM operated in the
frequency-shift mode. Section~III contains our measurement
results and the analysis of errors. In Sec.~IV the computational
results for the gradient of the Casimir force between two Ni
surfaces are presented. The calculation for the upper bound of
the magnetic interaction in our setup can be found in Sec.~V.
In Sec.~VI the reader will find the comparison between experiment
and theory for two magnetic test bodies. Section VII contains
the comparison with previously performed experiments.
Section~VIII is devoted to our conclusions and discussion.
Appendices A and B contain some details of mathematical calculations.

\section{Measurement scheme using dynamic AFM}

The dynamic AFM operated in the frequency-shift mode, used in
this experiment to measure the gradient of the Casimir force
between Ni-coated surfaces of a hollow glass sphere of
$R=61.71\pm 0.09\,\mu$m radius and a Si plate, is already
described in Refs.\cite{32,39} and  \cite{48}.
Here we present only a few main points necessary for understanding
 of the subsequent text and dwell only on details which were not
discussed previously.
The sphere was attached to the rectangular Si cantilever of an
AFM and the plate was mounted on top of a piezoelectric tube
capable of travelling a separation distance $z_{\rm piezo}$ of
$2.3\,\mu$m between the surfaces of a sphere and a plate.
The movement of the piezo was calibrated by a fiber
interferometer.
Both test bodies were cleaned using the special multi-step
cleaning procedure and placed in the vacuum chamber that was
capable of reaching a pressure of $10^{-9}\,$Torr by using
mechanical, turbo and ion pumps [see Fig.~1(a,b) in Ref.~\cite{39}
 for a layout of the setup].
The piezoelectric tube contained a small
magnet introduced by the piezotube
manufacturer which is not needed in
this experiment. The initial magnetic field was
measured to be $\approx 100\,$Gs using a Hall probe gaussmeter.
To prevent any effects from this field, we inserted a piece of
mu-metal magnetic shield between the top of the piezo tube and
the Ni-coated plate.
The residual magnetic field was below the detection resolution
of 0.1\,Gs. Both the initial and residual fields do not depend on
separation in the separation region
considered and do not contribute to the force gradient measured
in our work.

In a dynamic experiment using the frequency-shift mode the
measured
 quantity is the change of resonant frequency $\omega_0$ of
the periodically driven cantilever.\cite{33}
 The change of the resonant frequency from
$\omega_0$ to $\omega_r$ occurs under the influence of an external
force
\begin{equation}
F_{\rm tot}(a,T)=F_{\rm el}(a)+F(a,T),
\label{eq1}
\end{equation}
\noindent
acting between the sphere and the plate at the laboratory
temperature $T=300\,$K. Here $F_{\rm el}(a)$ is the electric
force caused by the voltages $V_i$ applied to the plate whereas
the sphere remains grounded and $F(a,T)$ is the Casimir force.
The absolute separation between the sphere and plate surface
is given by
\begin{equation}
a=z_{\rm piezo}+z_0,
\label{eq2}
\end{equation}
\noindent
where $z_0$ is the point of the closest approach between the
two surfaces, which is much larger than the separation on
contact in the dynamic experiments.
Note that even if $V_i=0$ there is some residual potential
difference $V_0$ between the sphere and the plate caused by
different connections and work functions of the polycrystalline
surface from patches and possible
adsorbates on their surfaces.

The shift of the resonance frequency of the cantilever
was detected by means of an optical interferometer.\cite{49,50}
To prevent any error in the sphere-plate separation $a$ due to
cantilever deflection under the influence of a force
$F_{\rm tot}(a,T)$, we have kept the interferometric cavity
length constant by means of an additional piezo, which was
controlled by a proportional-integral-derivative feedback
loop. Then the frequency shift
\begin{equation}
\Delta\omega(a)=\omega_r(a)-\omega_0,
\label{eq3}
\end{equation}
\noindent
was measured by the PLL frequency demodulator system
(here and below we omit an argument $T$ in the frequency shift
because $T$ is kept constant).
The output of the feedback loop provided by the PLL was the
resonant-frequency shift $\omega_r(a)-\omega_d$,
where $\omega_d$ is the set-point frequency of the PLL.
We made sure that at large separations above $2.2\,\mu$m
the frequency shift $\omega_r(a)-\omega_d$ remains constant
within the resolution limit. From this it follows that at
separations above $2.2\,\mu$m there is no influence of the
external force and  $\omega_r(a)=\omega_0$.
Finally the frequency shift (\ref{eq3}) was found from the two
measured quantities by the equation
\begin{equation}
\Delta\omega(a)=[\omega_r(a)-\omega_d]-(\omega_0-\omega_d).
\label{eq3a}
\end{equation}

In the linear regime, which holds for sufficiently small
oscillation amplitudes of the cantilever, the frequency shift
is given by\cite{39}
\begin{equation}
\Delta\omega(a)=-\frac{\omega_0}{2k}\,
\frac{\partial F_{\rm tot}(a,T)}{\partial a}
\equiv -\frac{\omega_0}{2k}F_{\rm tot}^{\prime}(a,T),
\label{eq4}
\end{equation}
\noindent
where $k$ is the spring constant of the cantilever
(maximum allowed amplitudes ensuring the applicability of the
linear regime are calculated in Ref.\cite{39}).

The electric force contributing to the total force (\ref{eq1})
in the configuration of a metal sphere above a metal plate can
be calculated precisely as\cite{9,51}
\begin{equation}
F_{\rm el}(a)=X(a,R)(V_i-V_0)^2.
\label{eq5}
\end{equation}
\noindent
Here the function $X(a,R)$ is given by
\begin{eqnarray}
&&
X(a,R)=2\pi\epsilon_0\sum_{n=1}^{\infty}
\frac{\coth\alpha-n\coth(n\alpha)}{\sinh(n\alpha)},
\nonumber\\
&&
\cosh\alpha=1+\frac{a}{R},
\label{eq6}
\end{eqnarray}
\noindent
where $\epsilon_0$ is the permittivity of the vacuum.
When using Eq.~(\ref{eq6}) in electrostatic calibrations
(see Sec.~III), it is convenient to present $X(a,R)$ as
the sum of powers\cite{9,52}
\begin{equation}
X(a,R)=-2\pi\epsilon_0\left[c_{-1}\frac{R}{a}+c_0+
c_1\frac{a}{R}+c_2\frac{a^2}{R^2}+\ldots\right],
\label{eq7}
\end{equation}
\noindent
where $c_1=0.5$, $c_0=-1.18260$, $c_1=22.2375$,
$c_2=-571.366$ etc.
Substituting Eqs.~(\ref{eq1}) and  (\ref{eq5}) in
Eq.~(\ref{eq4}), one can connect the measured frequency shift
with the gradient of the Casimir force
\begin{equation}
\Delta\omega(a)=-\beta(V_i-V_0)^2-
C\frac{\partial F(a,T)}{\partial a}.
\label{eq8}
\end{equation}
\noindent
Here $C\equiv\omega_0/(2k)$ and
$\beta\equiv\beta(z_0,z_{\rm piezo},C,R)=
C\partial X(a,R)/\partial a$.
Substituting Eq.~(\ref{eq7}) in the definition of $\beta$,
one obtains
\begin{equation}
\beta=\frac{\pi\epsilon_0RC}{a^2}\left(1-
2c_1\frac{a^2}{R^2}-4c_2\frac{a^3}{R^3}+\ldots\right),
\label{eq9}
\end{equation}
\noindent
where $a$ is expressed according to Eq.~(\ref{eq2}).

\section{Measurement results and error analysis for two
{\\}
magnetic bodies}

To find the gradient of the Casimir force as a function of
separation from the measured frequency shift by using
Eq.~(\ref{eq8}),
 one needs sufficiently precise values of the coefficients
$C$ and $\beta$, of the residual potential difference $V_0$,
and of the separation at the closest approach $z_0$.
These were found by means of electrostatic calibrations which
were performed using the dependence of the frequency shift on
the applied voltages in Eq.~(\ref{eq8}).
For this purpose 11 different voltages, $-64.5,\,\,-54.7,\,\,
-45.1,\,\,-35.3,\,\,-25.5,\,\,-17.7,\,\,-7.8,\,\,2.2,\,\,
12.5,\,\,22.3$ and 31.6\,mV were sequentially applied to the
plate while the sphere remained grounded.
With each applied voltage the plate was moved towards the sphere
starting at the maximum separation of $2.3\,\mu$m and the
frequency shift $\Delta\omega(a)$ was recorded at each 0.14\,nm.
To move the plate towards the sphere, continuous triangular
voltages at 0.01\,Hz were applied to the piezoelectric tube.
The small mechanical drift in the $z_{\rm piezo}$ was measured
to be 0.003\,nm/s and corrected using the procedure described
in Refs.\cite{32} and \cite{39}.
Measurement of the frequency shift $\Delta\omega(a)$ was repeated
three times with each applied voltage $V_i$. This resulted
in 33 measurement sets.

To perform the electrostatic calibration, the measured frequency
shift with a step of 1\,nm was found by interpolation. Then at
every 1\,nm $\Delta\omega$ was plotted as a function of the
applied voltage $V_i$ and the value of $V_0$ was identified as
the position of the parabola maximum.\cite{32,39}
The obtained values of $V_0$ as a function of separation are
plotted in Fig.~1 of Ref.\cite{48} over the separation region
from 220 to 1000\,nm.
As can be seen in this figure, $V_0$ does not depend on
separation indicating that the interacting regions of the
surfaces are clean or the adsorbed impurities are randomly
distributed with a submicrometer size scale and make
only a negligible
contribution to the total force.\cite{39}
The mean value of $V_0$ was found to be
$V_0=-17.7\pm 1\,$mV (here and below the errors are indicated
at a 67\% confidence level if another value is not stated
explicitly).

Next, we determined the coefficient $C$ and the separation at
the closest approach $z_0$ by fitting the data for the parabola
curvature $\beta$ to the theoretical expression in
Eq.~(\ref{eq9}).
A least $\chi^2$ fitting procedure was used which was repeated
by keeping the start points fixed at the closest separation
$z_0$, while the end point $z_{\rm end}$ measured from $z_0$
was varied from 150 to 1190\,nm.
In Fig.~\ref{fg1}(a) the obtained values of $C$ are seen to be
almost independent on the end point indicating the absence of
systematic errors from the calibration of $z_{\rm piezo}$,
mechanical drift etc. The obtained mean value is
$C=52.4\pm 0.16\,$kHz\,m/N.
In Fig.~\ref{fg1}(b) the respective values of $z_0$ are
presented as a function of $z_{\rm end}$.
They are also independent of $z_{\rm end}$ in the limits of
errors of the fitting procedure. The mean value is
$z_0=221.1\pm 0.4\,$nm.
Then the absolute separations $a$ between the sphere and the
plate are obtained from Eq.~(\ref{eq2}).
The error in the determination of the absolute separations,
$\Delta a$, is also equal to 0.4\,nm because the relative
separations, $z_{\rm piezo}$, are determined to a much higher
precision. We emphasize that our calibration parameters,
including absolute separations, are determined with
significantly smaller errors than it is common for
 sharp tips. The reason is that we use
a large perfectly shaped sphere made from the liquid phase
instead of rough surfaces where geometry is not known precisely.
Another specific feature of our experiment is that the
theoretical electric force in the sphere-plate geometry is
known exactly and the electric potential can be determined to
a high precision.

We are now in a position to find the gradients of the Casimir
force $F^{\prime}(a,T)\equiv\partial F(a,T)/\partial a$ from the
measured frequency shifts by using Eq.~(\ref{eq8}). They are again
found at each 0.14\,nm and then interpolated in order to get 33
values of the force gradient at each nanometer of the absolute
separation $a$ (starting from 223\,nm). We have checked the
statistical properties of the Casimir force gradient data obtained
in this way and made sure that they are characterized by a
Gaussian distribution (see Sec.~VII for more details).
 In Fig.~\ref{fg2} we plot as dots all 33 data
points for $F^{\prime}(a)$ with a step of 5nm starting from
the first integer separation 223\,nm, where our measurements were
performed. The solid line shows the mean values of the measured
gradients of the Casimir force found from 33 measurements.
In the inset the same information is presented over a more narrow
separation region which gives the possibility to demonstrate all
the data points with a step of 1\,nm.

In this experiment  the mean gradients of the
Casimir force are burdened by  errors of two types, the random
and the systematic.
The total experimental error is obtained as a combination of these two
taking into account their distribution laws (see Refs.\cite{9} and \cite{10}
for details).
The random error $\Delta^{\! r}F^{\prime}(a)$
calculated from 33 repetitions at a 67\% confidence level using
the Student distribution [the Student coefficient
$t_{(1+0.67)/2}(32)=1$] as a function of separation is shown by the
short-dashed line in Fig.~\ref{fg3}. As can be seen in the figure,
$\Delta^{\! r}F^{\prime}(a)$ does not depend on separation.
The systematic error is determined by the instrumental noise
including the background noise level, by the errors in
calibration, and by the errors in the gradient of the subtracted
electrostatic force. Taking into account that all these errors
are characterised by Gaussian distributions, to obtain the total
systematic error $\Delta^{\! s}F^{\prime}(a)$ they were
combined in quadrature.
 The obtained values of  $\Delta^{\! s}F^{\prime}(a)$
 at a 67\% confidence level, as a function of separation,
are shown  in Fig.~\ref{fg3} by the
long-dashed line. The increase of  $\Delta^{\! s}F^{\prime}(a)$
at shorter separations is caused by the errors in the subtracted
electrostatic force.
As can be seen in Fig.~\ref{fg3}, the systematic error is
from a factor of 6 to a factor of 4 larger than the random error,
as is typical for precise experiments of metrological quality.
The total experimental error
$\Delta^{\! t}F^{\prime}(a)$  at a 67\% confidence level is
obtained in quadrature from the random and systematic errors.
It is shown by the solid line in Fig.~\ref{fg3}.
One can see that the total experimental error at all separations
is mostly determined by the systematic error.

\section{Calculation of the Casimir interaction between two
${\rm\bf Ni}$ bodies}

Here we calculate the gradient of the Casimir force in the
experimental configuration of a Ni-coated sphere and a Ni-coated
plate. Given the thicknesses of Ni coatings
($d_1=250\pm 1\,$nm and $d_2=210\pm 1\,$nm on the plate and the
sphere, respectively), one can consider them as a solid Ni
ball near a Ni semispace.\cite{9}
Using the PFA, the gradient of the Casimir force is given by
\begin{equation}
F_{\rm PFA}^{\prime}(a,T)=2\pi R{\cal F}_{pp}^{\prime}(a,T),
\label{eq10}
\end{equation}
\noindent
where ${\cal F}_{pp}(a,T)$ is the free energy of the Casimir
interaction per unit area of two parallel Ni semispaces
spaced $a$ nanometers apart in thermal equilibrium at
temperature $T$. According to the Lifshitz theory,\cite{3,18}
${\cal F}_{pp}(a,T)$ can be presented as the sum from $l=0$
to $l=\infty$ over the Matsubara frequencies $\xi_l$
(see Sec.~I). Then the gradient of the Casimir force (\ref{eq10})
takes the form\cite{39,48}
\begin{equation}
F_{\rm PFA}^{\prime}(a,T)=2k_BTR
\sum_{l=0}^{\infty}{\vphantom{\sum}}^{\prime}
\int_{0}^{\infty}q_lk_{\bot}dk_{\bot}\sum_{\alpha}
\frac{r_{\alpha}^2}{e^{2aq_l}-r_{\alpha}^2}.
\label{eq11}
\end{equation}
\noindent
Here, $q_l^2=k_{\bot}^2+\xi_l^2/c^2$, $k_{\bot}$ is the
projection of the wave vector on the plate, and the prime
following the summation sign multiplies the term with $l=0$ by
1/2. The index $\alpha$ takes the two values TM and TE and
denotes the transverse magnetic and transverse electric
polarizations of the electromagnetic field. The respective
reflection coefficients $r_{\alpha}$ calculated along the
imaginary frequency axis have the following explicit form:
\begin{eqnarray}
&&
r_{\rm TM}\equiv r_{\rm TM}(i\xi_l,k_{\bot})=
\frac{\varepsilon(i\xi_l)q_l-k_l}{\varepsilon(i\xi_l)q_l+k_l},
\nonumber \\
&&
r_{\rm TE}\equiv r_{\rm TE}(i\xi_l,k_{\bot})=
\frac{\mu(i\xi_l)q_l-k_l}{\mu(i\xi_l)q_l+k_l},
\label{eq12} \\
&&
k_l=\left[k_{\bot}^2+\varepsilon(i\xi_l)\mu(i\xi_l)
\frac{\xi_l^2}{c^2}\right]^{1/2}.
\nonumber
\end{eqnarray}
\noindent
The main properties of the magnetic permeability $\mu$ of a
boundary material (Ni) calculated at the imaginary Matsubara
frequencies are discussed in Sec.~I.

To apply Eqs.~(\ref{eq11}) and (\ref{eq12}) for the calculation
of the Casimir interaction, one needs to have the values of
$\varepsilon(i\xi_l)$ up to sufficiently large values of $l$
and $\mu(0)$ (see Sec.~I).
The dielectric permittivity at the imaginary Matsubara
frequencies is obtained by means of the Kramers-Kronig relation
from ${\rm Im}\,\varepsilon(\omega)=2n_1(\omega)n_2(\omega)$,
where $n_1(\omega)$ and $n_2(\omega)$ are the real and imaginary
parts of the complex index of refraction, respectively, measured
and tabulated over a wide frequency region.\cite{53}
An application of the Kramers-Kronig relation requires, however,
the optical data at much lower frequencies than it may
become available in any foreseeable future. Because of this,
the problem arises on how to extrapolate the data for
${\rm Im}\,\varepsilon(\omega)$ to lower frequencies down to zero
frequency. In Sec.~I two approaches to the resolution of
this problem proposed in the literature are mentioned.
According to the first approach, which seems to be the most
natural and straightforward from a theoretical point of view,
in any extrapolation the properties of boundary materials should
be described as precise as possible. Specifically, the relaxation
properties of conduction electrons at low frequencies should be
taken into account by means of the commonly accepted Drude
dielectric function
\begin{equation}
\varepsilon_D(\omega)=1-\frac{\omega_p^2}{\omega[\omega+i\gamma(T)]},
\label{eq13}
\end{equation}
\noindent
where $\omega_p$ is the plasma frequency and $\gamma(T)$ is the
relaxation parameter. This approach was called the
{\it Drude model approach}. The theoretical predictions for the
Casimir interaction obtained in this way
were excluded by several experiments with
metallic test bodies\cite{9,10,39,42,43,44,45,48}
performed by R.\  S.\  Decca and U.\  Mohideen groups.
At the same time in two other experiments\cite{54,55}
of S.\ K.\ Lamoreaux group the Drude
model approach was claimed to be in agreement with the data.
This conclusion, however, has been
questioned.\cite{56,57,58,59}
Furthermore, several experiments performed by
E.\ A.\ Cornell and U.\ Mohideen groups
 with dielectric materials turned
out to be in contradiction with theoretical predictions if the
free charge carriers are included in the Lifshitz
theory.\cite{9,10,60,61,62,63,64,65}
Besides, the inclusion of the relaxation properties of charge
carriers or taking into account the free charge carriers for
dielectrics in the Lifshitz theory were found to violate
the third law of thermodynamics (the Nernst heat
theorem).\cite{6,10,66,67}
This, however, happens at zero temperature and is not
directly relevant to any experimental work.

The second proposed approach suggested to extrapolate the optical
data for metals to zero frequency by means of the plasma model
$\varepsilon_p(\omega)$. The latter is obtained from
Eq.~(\ref{eq13})
by putting $\gamma(T)=0$, i.e., by disregarding the relaxation
properties of free charge carriers (note that for permittivities
having the second order pole at zero frequency the Kramers-Kronig
relation is modified accordingly\cite{68}).
The {\it plasma model approach} was found to be consistent with
the measurement data of the experiments with
metallic test bodies.\cite{9,10,39,42,43,44,45,48}
The Drude and the plasma model approaches are the subject of
continuing discussions in the literature.\cite{69,70,71,72}
Below we perform computations using both approaches on equal
terms and compare the obtained results between themselves and
with the measurement data.

We obtained the dielectric permittivity $\varepsilon(i\xi_l)$
from the optical data\cite{53} for the complex index of
 refraction of Ni using the Kramers-Kronig relation.
 The data were first extrapolated to zero frequency by using
 either the Drude or the plasma models. In so doing we have used
the plasma frequency of Ni $\omega_p=4.89\,$eV and the relaxation
parameter at $T=300\,$K $\gamma=0.0436\,$eV according to
Refs,\cite{53} and \cite{73}.
Our Ni-coated test bodies did not possess a spontaneous
magnetization due to sufficiently thick coatings and weak
environment magnetic fields. The magnetic properties of Ni were
described by a static magnetic permeability $\mu(0)=110$.
For all Matsubara frequencies with $l\geq 1$ at $T=300\,$K it
holds $\mu(i\xi_l)=1$ because $\mu(i\xi)$ rapidly falls to unity
with increasing $\xi$ (see Sec.~I).

Equation (\ref{eq11}) was obtained using the PFA and, thus, is
not exact. Recently the gradient of the Casimir force in a
sphere-plate configuration was calculated exactly and the
corrections to the PFA result were found.\cite{74,75,76,77}
According to these papers the exact force gradient between the
sphere of large radius and the plate is equal to
\begin{eqnarray}
F^{\prime}(a,T)&=&F_{\rm PFA}^{\prime}(a,T)
\left[1+\delta_{\rm corr}^{\rm PFA}(a,T,R)\right]
\nonumber \\[1mm]
&=&F_{\rm PFA}^{\prime}(a,T)
\left[1+\theta(a,T)\frac{a}{R}+o\left(\frac{a}{R}\right)\right],
\label{eq14}
\end{eqnarray}
\noindent
where $F_{\rm PFA}^{\prime}$ is given in Eq.~(\ref{eq11}).
In Ref.\cite{76} the coefficient $\theta$, as a function of
separation, was calculated for Au at both $T=0$ and $T=300\,$K
using the Drude model approach. In the separation region from
220 to 550\,nm the obtained results at $T=300\,$K only slightly
differ from those for ideal metal surfaces
considered at $T=300\,$K in the framework of thermal quantum
field theory.
This demonstrates a very weak dependence of $\theta$
on the plasma frequency
$\omega_p$, relaxation parameter $\gamma$ and optical data
within this separation region.
 Because of this, one can use the values of $\theta$
found in Ref.\cite{76} for Ni as well.
It was also shown\cite{77} that at $T=0$ the values of $\theta$
calculated using the plasma model are sandwiched between those
calculated using the Drude model and for ideal metal surfaces.
This allows one to approximate the values of $\theta$
at $T=300\,$K in the plasma model approach by those
for ideal metals at the same temperature. In Fig.~\ref{fg4} we
present by the upper and lower solid lines the correction to
PFA $\delta_{\rm corr}$ at $T=300\,$K in percent for the Drude
and plasma model approaches, respectively (note that the
correction of order $a^2/R^2\sim 0.3\times 10^{-4}\,$
can be neglected).
As can be seen in Fig.~\ref{fg4}, the error from using the PFA
is substantially smaller than $a/R$. The latter value for
this error was used
previously.\cite{9,10,39,42,43,44,45,48,60,61,64,65}
Thus, the analysis of previous experiments was highly
conservative.

One more correction factor which should be introduced in
Eq.~(\ref{eq11}) is due to the surface roughness.
The root-mean-square roughness on the sphere and the plate
was investigated by means of an AFM with a sharp tip and found
to be $\delta_s=1.5\,$nm and $\delta_p=1.4\,$nm,
respectively. For so small $a$ roughness at separations above
200\,nm one can use the multiplicative approach.\cite{9,10}
In the framework of this approach the force gradient with
account of surface roughness is given by\cite{9,10}
\begin{eqnarray}
&&
F_{R}^{\prime}(a,T)=F^{\prime}(a,T)
\left[1+\delta_{\rm corr}^{R}(a)\right],
\label{eq15} \\[2mm]
&&
\delta_{\rm corr}^{R}(a)=10\frac{\delta_s^2+\delta_p^2}{a^2}+
105\frac{(\delta_s^2+\delta_p^2)^2}{a^4}.
\nonumber
\end{eqnarray}
\noindent
In Fig.~\ref{fg4} the correction due to surface roughness
$\delta_{\rm corr}^{R}$ in percent is shown by the dashed line as a
function of separation. As can be seen in Fig.~\ref{fg4},
the corrections due to deviations from the PFA and due to
surface roughness are of opposite signs and give only minor
contributions to the force gradient.

Now we are in a position to calculate the gradient of the
Casimir force $F^{\prime}(a,T)$ between two Ni surfaces at
$T=300\,$K with account of all correction factors.
Computations were performed by Eqs.~(\ref{eq11}), (\ref{eq12}),
(\ref{eq14}), and (\ref{eq15}), using the Drude
[$F_{R,D}^{\prime}(a,T)$] and plasma [$F_{R,p}^{\prime}(a,T)$] model
approaches. The computational results are shown in Fig.~\ref{fg5}
by the upper and lower lines, respectively.
We emphasize that for two Ni test bodies
$F_{R,D}^{\prime}> F_{R,p}^{\prime}$ at all separations.
This is quite the reverse to the case of two Au test bodies and
leads to important consequences discussed in Sec.~VII.
To obtain a striking understanding of the difference between
the predictions of the two approaches, in Fig.~\ref{fg6}(a) we
also plot the difference
\begin{equation}
F_{\!\rm diff}^{\prime}(a,T)\equiv
F_{R,D}^{\prime}(a,T)-F_{R,p}^{\prime}(a,T)
\label{eq16}
\end{equation}
at $T=300\,$K as a function of separation.
In the same figure the dashed line reproduces from
Fig.~\ref{fg3} the total experimental error in measurements
of force gradients.
As can be seen in Fig.~\ref{fg6}(a), $F_{\!\rm diff}^{\prime}$
is well above the total experimental error
determined at a 67\% confidence level up to almost 450\,nm.
However, with increasing separation distance, the
magnitudes of $F_{\!\rm diff}^{\prime}$ fall below the total
experimental error. In Fig.~\ref{fg6}(b) we also plot
as a function of separation the relative difference (in percent)
between the predictions of two theoretical approaches
$F_{\!\rm diff}^{\prime}/F_{R,p}^{\prime}$.
It can be seen that the relative difference for Ni test bodies
increases with increasing
separation and achieves 10\% at separations
above 500\,nm (see Sec.~VII where the cases of Ni and Au test
bodies are compared).

\section{Calculation of magnetic interaction in the experimental
setup}

Before the measurement data could be compared with the above
computational results for the gradient of the Casimir force
using different theoretical approaches, due attention should be
focused on magnetic interactions. Note that in our
experiment both interacting surfaces are ferromagnetic films and
consist of many domains. Because of this, it is necessary to
calculate the maximum possible contribution of magnetic forces into
the measurement results. First, we calculate the energy of
magnetic interaction per unit area of two plane parallel films.
For this purpose the domain structure of the films of sizes
$L_x^{(1)}\times L_y^{(1)}$ and $L_x^{(2)}\times L_y^{(2)}$ is
periodically continued for infinite planes.
Then, using the PFA, we calculate the upper bound for the
gradient of the magnetic force acting between a sphere and a plate
coated with magnetic films.
Note that PFA in the form of Eq.~(\ref{eq10}) is  applicable not
only to  Casimir forces, but, for instance, to the electric forces
which decrease with separation less rapidly.\cite{9}
It was shown\cite{grav} that a more general formulation of the PFA
(the so-called Derjaguin method\cite{Der}) is applicable even to
volumetric forces which do not decrease with separation. In our case of
magnetic forces we have checked that the PFA in both formulations
leads to coincident results.

The magnetic field created by a magnetic body $V_1$ at the point
$(x_2,y_2,z_2)$ is given by the expression\cite{15}
\begin{equation}
\mbox{\boldmath$H$}(x_2,y_2,z_2)=\int_{V_1}dx_1dy_1dz_1
\frac{3(\mbox{\boldmath$n$}_r\cdot\mbox{\boldmath$M$}_1)\mbox{\boldmath$n$}_r
-\mbox{\boldmath$M$}_1}{|\mbox{\boldmath$r$}|^3}.
\label{eq17}
\end{equation}
\noindent
Here, the integration is extended over the coordinates $(x_1,y_1,z_1)$
of the body $V_1$, the unit vector $\mbox{\boldmath$n$}_r$ is
given by
\begin{equation}
\mbox{\boldmath$n$}_r=\frac{\mbox{\boldmath$r$}}{|\mbox{\boldmath$r$}|},
\label{eq18}
\end{equation}
\noindent
the vector {\boldmath$r$} is directed from the point
$(x_1,y_1,z_1)$
to the point $(x_2,y_2,z_2)$ and
\begin{equation}
|\mbox{\boldmath$r$}|=\left[(x_2-x_1)^2+(y_2-y_1)^2+(z_2-z_1)^2
\right]^{1/2}.
\label{eq19}
\end{equation}
\noindent
The magnetization distribution
$\mbox{\boldmath$M$}_1\equiv\mbox{\boldmath$M$}_1(x_1,y_1,z_1)$
of the body $V_1$ depends on a point.
The energy of the magnetic interaction between the bodies $V_1$
and $V_2$ is given by\cite{15}
\begin{equation}
E_m=-\int_{V_2}dx_2dy_2dz_2
\mbox{\boldmath$M$}_2(x_2,y_2,z_2)\cdot
\mbox{\boldmath$H$}(x_2,y_2,z_2),
\label{eq20}
\end{equation}
\noindent
where
$\mbox{\boldmath$M$}_2\equiv\mbox{\boldmath$M$}_2(x_2,y_2,z_2)$
is the magnetic distribution of the second body.

The orientation of magnetization of separate domains in the
magnetic films depends on the film thicknesses. Thus, one can
obtain in-plane magnetization only in very thin films.
With increasing film thickness up to 150\,nm and more, the easy
direction is out-of-plane perpendicular to it.\cite{78,79,80}
Although in our experiment the film thicknesses satisfy this
condition, below we calculate the upper bounds of the magnetic
interaction for both out-of-plane and in-plane magnetizations
and show that in both cases the gradient of magnetic force  is
negligibly small.
Note that any other alignment of domains (which cannot occur in thin
films but might be possible in thick magnetic bodies) can be presented
as a superposition of these two. The results obtained below concerning
the smallness of the magnetic interaction are valid for films consisting of
many domains. The numerical estimations use the domain sizes, as in
our experiment.
 We start from the most realistic case of an
out-of-plane magnetization.

\subsection{Out-of-plane magnetization}

For magnetization perpendicular to the plane of first and second
films one has
\begin{equation}
\mbox{\boldmath$M$}_{1,2}=(0,0,M_z^{(1,2)}),\quad
M_z^{(1,2)}\equiv M_z^{(1,2)}(x_{1,2},y_{1,2}),
\label{eq21}
\end{equation}
\noindent
i.e., magnetizations do not vary with film surface distances.
We first assume that the magnetizations
$M_z^{(1,2)}$ take the values $M_s$ and $-M_s$ with equal
probability (it is known that\cite{78}
$M_s=435\,\mbox{emu/cm}^3$). Here it is assumed that the films
do not possess a spontaneous magnetization as predicted by
experimental conditions (the case of films possessing a spontaneous
magnetization is considered next).
The magnetic interaction between the two parallel films of finite
area consisting of randomly distributed domains can be calculated
using the forma\-lism developed earlier\cite{81} to take into
account the impact of surface roughness to the Casimir force.
For this purpose we perform the periodic
continuation of the functions
$M_z^{(1,2)}(x_{1,2},y_{1,2})$ as odd functions with periods
$2L_x^{(1,2)}$ and $2L_y^{(1,2)}$ over the whole planes $(x_1,y_1)$ and
$(x_2,y_2)$, respectively. The obtained periodic functions can
be expanded in the Fourier series
\begin{equation}
M_z^{(1,2)}(x_{1,2},y_{1,2})=\sum_{k,n=1}^{\infty}
M_{kn}^{(1,2)}\sin\frac{k\pi x_{1,2}}{L_x^{(1,2)}}
\sin\frac{n\pi y_{1,2}}{L_y^{(1,2)}},
\label{eq22}
\end{equation}
\noindent
where $M_{kn}^{(1,2)}$ are the Fourier coefficients.

Now one can use the standard formalism of magnetic force
microscopy\cite{82,83} to calculate the magnetic energy
(\ref{eq20})
between the two parallel films spaced at a separation $a$.
We emphasize, however, that when scanning a sharp tip above
the boundary of two neighboring
domains in magnetic force microscopy,
they are usually modeled by a periodic structure.\cite{82,84}
In this case the quantities $L_{x,y}^{(1,2)}$ in Eq.~(\ref{eq22})
are replaced with the characteristic sizes of the magnetic domains
 $D_{x,y}^{(1,2)}$. For randomly distributed domains, as in
our case, the dominant contribution to the right-hand side of
Eq.~(\ref{eq22}) is given by the item numbered with rather large
indices
$k\approx L_x^{(1,2)}/D_x^{(1,2)}$ and
$n\approx L_y^{(1,2)}/D_y^{(1,2)}$.
Thus, taking into account that the size of magnetic domains is
approximately equal to the film thickness, we obtain for the
domains on the first film
$D_x^{(1)}\approx D_y^{(1)}\approx d_1=250\,$nm.
In  a similar way
$D_x^{(2)}\approx D_y^{(2)}\approx d_2=210\,$nm.
Then, using the sizes of the first film $L_x^{(1)}=0.9\,$cm
and $L_y^{(1)}=1.1\,$cm, we arrive at $k\approx 3.6\times 10^4$
and $n\approx 4.4\times 10^4$.
After calculations using Eqs.~(\ref{eq17})--(\ref{eq22}), for
the magnetic energy per unit area of the first film one arrives
at (see Appendix A for details)
\begin{eqnarray}
&&
\frac{E_m(a)}{L_x^{(1)}L_y^{(1)}}=\frac{1}{L_x^{(1)}L_y^{(1)}}
\sum_{k,n=1}^{\infty}\sum_{k^{\prime},n^{\prime}=1}^{\infty}
M_{kn}^{(1)}M_{k^{\prime}n^{\prime}}^{(2)}
X_{kk^{\prime}}Y_{nn^{\prime}}
\label{eq23} \\
&&
~~~
\times\int_{a}^{a+d_2}dz_2\left[(z_2+d_1)\Phi_{kn}(z_2+d_1)-
z_2\Phi_{kn}(z_2)\right].
\nonumber
\end{eqnarray}
\noindent
Here, the functions $X_{kk^{\prime}}$ and $Y_{nn^{\prime}}$
are defined as
\begin{eqnarray}
&&
X_{kk^{\prime}}=\int_0^{L_x^{(1)}}dx
\sin\frac{\pi kx}{L_x^{(1)}}\sin\frac{\pi k^{\prime}x}{L_x^{(2)}},
\nonumber\\[-1mm]
&&
\label{eq24} \\
&&
Y_{nn^{\prime}}=\int_0^{L_y^{(1)}}dy
\sin\frac{\pi ny}{L_y^{(1)}}\sin\frac{\pi n^{\prime}y}{L_y^{(2)}}
\nonumber
\end{eqnarray}
\noindent
and the function $\Phi_{kn}(z)$ is defined by
\begin{equation}
\Phi_{kn}(z)=\frac{2\pi}{z}e^{-\gamma_{kn}z}, \quad
\gamma_{kn}=\pi\left[\left(\frac{k}{L_x^{(1)}}\right)^2+
\left(\frac{n}{L_y^{(1)}}\right)^2\right]^{1/2}.
\label{eq25}
\end{equation}

Using the PFA in Eq.~(\ref{eq10}), one can obtain the gradient of
the magnetic force acting between a sphere and a plate.
For this purpose we replace ${\cal F}_{pp}$ in Eq.~(\ref{eq10})
with $E_m/(L_x^{(1)}L_y^{(1)})$ defined in Eq.~(\ref{eq23}) and
perform differentiation with respect to $a$. The result is
\begin{eqnarray}
&&
F_m^{\prime}(a)=\frac{4\pi^2R}{L_x^{(1)}L_y^{(1)}}
\sum_{k,n=1}^{\infty}\sum_{k^{\prime},n^{\prime}=1}^{\infty}
M_{kn}^{(1)}M_{k^{\prime}n^{\prime}}^{(2)}
e^{-\gamma_{kn}a}
\label{eq26} \\
&&
~~~
\times
(1-e^{-\gamma_{kn}d_1})(1-e^{-\gamma_{kn}d_2})
X_{kk^{\prime}}Y_{nn^{\prime}}.
\nonumber
\end{eqnarray}
\noindent
Note that the factors $X_{kk^{\prime}}$ and $Y_{nn^{\prime}}$
do not average to zero due to the finitness of
$L_x^{(1)}$, $L_y^{(1)}$ and
$L_x^{(2)}\approx L_y^{(2)}\sim 2R$, i.e., due to the
boundary effects.

Now we estimate the gradient of the magnetic force (\ref{eq26})
by considering the dominant contribution to Eq.~(\ref{eq26}).
As discussed above, the dominant contribution is given by
$k\approx L_x^{(1)}/D_x^{(1)}$ and
$k^{\prime}\approx L_x^{(2)}/D_x^{(2)}$ and hence by
\begin{eqnarray}
&&
X_{kk^{\prime}}=\int_{0}^{L_x^{(1)}}\!\!\!dx
\sin\frac{\pi x}{D_x^{(1)}}
\sin\frac{\pi x}{D_x^{(2)}}
\label{eq27} \\
&&~~
=\frac{D_x^{(1)}D_x^{(2)}}{2\pi}\left\{
\frac{1}{D_x^{(2)}-D_x^{(1)}}\left[
\sin\frac{\pi L_x^{(1)}}{D_x^{(1)}}
\cos\frac{\pi L_x^{(1)}}{D_x^{(2)}}-
\cos\frac{\pi L_x^{(1)}}{D_x^{(1)}}
\sin\frac{\pi L_x^{(1)}}{D_x^{(2)}}\right]
\right.
\nonumber \\
&&~~~~~
\left.
-\frac{1}{D_x^{(1)}+D_x^{(2)}}\left[
\sin\frac{\pi L_x^{(1)}}{D_x^{(1)}}
\cos\frac{\pi L_x^{(1)}}{D_x^{(2)}}+
\cos\frac{\pi L_x^{(1)}}{D_x^{(1)}}
\sin\frac{\pi L_x^{(1)}}{D_x^{(2)}}\right]
\right\}.
\nonumber
\end{eqnarray}
\noindent
Taking into account, that $L_x^{(1)}/D_x^{(1)}$ is an integer
number, one obtains
\begin{eqnarray}
&&
|X_{kk^{\prime}}|\leq\frac{D_x^{(1)}D_x^{(2)}}{2\pi}
\left(
\frac{1}{D_x^{(1)}-D_x^{(2)}}
-\frac{1}{D_x^{(1)}+D_x^{(2)}}\right)
\nonumber\\
&&~~~
=\frac{D_x^{(1)}}{\pi\left[\left(\frac{D_x^{(1)}}{D_x^{(2)}}
\right)^2-1\right]}\approx\frac{D_x^{(1)}}{0.4\pi}
\lesssim D_x^{(1)}.
\label{eq28}
\end{eqnarray}
\noindent
In a similar way for $n\approx L_y^{(1)}/D_y^{(1)}$,
$n^{\prime}\approx L_y^{(2)}/D_y^{(2)}$ we get
\begin{equation}
|Y_{nn^{\prime}}|\lesssim D_y^{(1)}.
\label{eq29}
\end{equation}

Using Eqs.~(\ref{eq28}) and (\ref{eq29}) we calculate the
dominant contribution to the gradient of magnetic force
({\ref{eq26}) at different separations.
Thus, at $a=223$, 250 and 300\,nm its magnitude is equal
to $1.4\times 10^{-3}$, $8.6\times 10^{-4}$ and
$3.5\times 10^{-4}\,\mu$N/m, respectively.
The gradient of the magnetic force further decreases in
magnitude with increasing separation. Numerical computations
using Eq.~(\ref{eq26}) show that at all separations the
total magnitude of the gradient of the magnetic forces due to
randomly distributed domains
$|F_m^{\prime}(a)|<10^{-2}\,\mu$N/m, i.e., much smaller than the
total error in the dynamic measurements of the Casimir interaction
(see Sec.~III).

The above calculations were performed under an assumption that
there is no spontaneous magnetization in our Ni films.
Now we include the case that there is some excess in the magnetization of
domains in one direction.
 This can be described by adding a nonzero  term
$M_{00}^{(1,2)}$ on the right-hand side of Eq.~(\ref{eq22}) for
$M_{z}^{(1,2)}$. The magnetic energy per unit area of two
parallel discs of $L_x^{(1)}/2$ radii arising due to such term
is obtained as (see Appendix A for details)
\begin{equation}
E_{sm}(a)= 4\pi M_{00}^{(1)}M_{00}^{(2)}
\int_{a}^{a+d_2}\!\!\!dz_2
\left[
\frac{z_2}{\sqrt{(L_x^{(1)})^2+4z_2^2}}-
\frac{z_2+d_1}{\sqrt{(L_x^{(1)})^2+4(z_2+d_1)^2}}
\right].
\label{eq30}
\end{equation}
\noindent
Then, the gradient of the magnetic force due to the spontaneous
magnetization is found by using the PFA in Eq.~(\ref{eq10}) where
we replace ${\cal F}_{pp}$ with $E_{sm}$ defined in
Eq.~(\ref{eq30}).
The result is
\begin{eqnarray}
&&
F_{sm}^{\prime}(a)= 8\pi^2R M_{00}^{(1)}M_{00}^{(2)}
\left[
\frac{a+d_2}{\sqrt{(L_x^{(1)})^2+4(a+d_2)^2}}-
\frac{a}{\sqrt{(L_x^{(1)})^2+4a^2}}
\right.
\nonumber \\
&&~~~~~~
\left.
-\frac{a+d_1+d_2}{\sqrt{(L_x^{(1)})^2+4(a+d_1+d_2)^2}}+
\frac{a+d_1}{\sqrt{(L_x^{(1)})^2+4(a+d_1)^2}}
\right].
\label{eq31}
\end{eqnarray}
\noindent
We calculate the quantity (\ref{eq31}) in an extreme case when
the magnetic moments of all domains are directed in one direction.
In this case $|M_{00}^{(1)}|=|M_{00}^{(2)}|=M_s$.
Then calculations using Eq.~(\ref{eq31}) result in
$|F_{sm}^{\prime}|\approx 2.6\times 10^{-5}\,\mu$N/m at
$a=223\,$nm
and even smaller values at larger separations.
Such small magnitudes for the gradient of the magnetic force due to
spontaneous magnetization are explained by the fact that in the
considered separation region this force depends on separation
only slightly. Thus, for out-of-plane magnetization of Ni films
one can neglect any influence of the magnetic interaction when
measuring the gradient of the Casimir force.

\subsection{In-plane magnetization}

Now we consider the magnetic interaction between Ni-coated
surfaces of a plate and a sphere under the assumption that
separate domains are characterized by the in-plane magnetization
(as discussed above, this might happen for sufficiently thin magnetic
films).

For the in-plane magnetization, one can choose the coordinate
system in such a way that for the first film
\begin{equation}
\mbox{\boldmath$M$}_{1}=(M_x^{(1)},0,0),\quad
M_x^{(1)}\equiv M_x^{(1)}(x_{1},y_{1}).
\label{eq32}
\end{equation}
\noindent
We further assume that there is no spontaneous magnetization
so that $M_x^{(1)}=\pm M_s$ with equal probability.
The in-plane magnetization of the second film may make an
angle $\alpha$ with the $x$ axis. Because of this
\begin{equation}
\mbox{\boldmath$M$}_{2}=(M_x^{(2)},M_y^{(2)},0),
\label{eq33}
\end{equation}
\noindent
where both components depend on the position and take random
values
\begin{equation}
M_x^{(2)}=\pm M_s\cos\alpha, \quad
M_y^{(2)}=\pm M_s\sin\alpha.
\label{eq34}
\end{equation}

Similar to Sec.~VA, we extrapolate the quantities
$M_x^{(1)}(x_1,y_1)$
and  $M_{x,y}^{(2)}(x_2,y_2)$ to the entire planes $(x_1,y_1)$ and
$(x_2,y_2)$  as odd functions
with the periods $2L_x^{(1,2)}$ and $2L_y^{(1,2)}$,
respectively. The obtained periodic functions can be expanded
in the Fourier series
\begin{eqnarray}
&&
M_x^{(1)}(x_{1},y_{1})=\sum_{k,n=1}^{\infty}
\tilde{M}_{kn}^{(1)}\sin\frac{k\pi x_{1}}{L_x^{(1)}}
\sin\frac{n\pi y_{1}}{L_y^{(1)}},
\nonumber \\
&&
M_{x,y}^{(2)}(x_{2},y_{2})=\sum_{k,n=1}^{\infty}
\tilde{M}_{x,y;kn}^{(2)}\sin\frac{k\pi x_{2}}{L_x^{(2)}}
\sin\frac{n\pi y_{2}}{L_y^{(2)}}.
\label{eq35}
\end{eqnarray}
\noindent
After caclulations using Eqs.~(\ref{eq17})--(\ref{eq20}) and
(\ref{eq32})--(\ref{eq35}) one obtains an expression for the
magnetic energy per unit area of two parallel films
(see Appendix B for details)
\begin{eqnarray}
&&
\frac{E_m(a)}{L_x^{(1)}L_y^{(1)}}=\frac{2\pi}{L_x^{(1)}L_y^{(1)}}
\sum_{k,n=1}^{\infty}\sum_{k^{\prime},n^{\prime}=1}^{\infty}
\tilde{M}_{kn}^{(1)}\left\{
\vphantom{\frac{kn\pi^2}{L_x^{(1)})L_y^{(1)}\gamma_{kn}^2}}
\tilde{M}_{x;k^{\prime}n^{\prime}}^{(2)}
X_{kk^{\prime}}Y_{nn^{\prime}}
\right.
\nonumber \\
&&~~
\times\int_{a}^{a+d_2}\!\!dz_2\left[
-\frac{k^2\pi^2}{(L_x^{(1)})^2\gamma_{kn}^2}e^{-\gamma_{kn}z_2}
\left(1-e^{-\gamma_{kn}d_1}\right)\right.
\nonumber \\
&&~~~~~~~~~
\left.
\vphantom{\frac{k^2\pi^2}{(L_x^{(1)})^2\gamma_{kn}^2}}
-{\rm Ei}(-\gamma_{kn}z_2)+{\rm Ei}[-\gamma_{kn(}z_2+d_1)]
\right]
\label{eq36} \\
&&
~~~\left.
+\tilde{M}_{y;k^{\prime}n^{\prime}}^{(2)}
\tilde{X}_{kk^{\prime}}\tilde{Y}_{nn^{\prime}}
\frac{kn\pi^2}{L_x^{(1)}L_y^{(1)}\gamma_{kn}^2}
\left(1-e^{-\gamma_{kn}d_1}\right)
\int_{a}^{a+d_2}\!\!dz_2e^{-\gamma_{kn}z_2}
\right\}.
\nonumber
\end{eqnarray}
\noindent
Here, the quantities $X_{kk^{\prime}}$ and $Y_{nn^{\prime}}$
are defined in Eq.~(\ref{eq24}), $\gamma_{kn}$ is
defined in Eq.~(\ref{eq25}), ${\rm Ei}(t)$ is the
exponential integral
and
\begin{eqnarray}
&&
\tilde{X}_{kk^{\prime}}=\int_0^{L_x^{(1)}\!\!}dx
\sin\frac{\pi kx}{L_x^{(1)}}\cos\frac{\pi k^{\prime}x}{L_x^{(2)}},
\nonumber\\[-1mm]
&&
\label{eq37} \\
&&
\tilde{Y}_{nn^{\prime}}=\int_0^{L_y^{(1)}}\!\!dy
\sin\frac{\pi ny}{L_y^{(1)}}\cos\frac{\pi n^{\prime}y}{L_y^{(2)}}.
\nonumber
\end{eqnarray}

The gradient of the magnetic force between a sphere and a plate
is obtained from the PFA in Eq.~(\ref{eq10}) by replacing
${\cal F}_{pp}$ with the magnetic energy per unit area defined
in Eq.~({\ref{eq36}). This leads us to the following result:
\begin{eqnarray}
&&
F_m^{\prime}(a)=\frac{4\pi^2R}{L_x^{(1)}L_y^{(1)}}
\sum_{k,n=1}^{\infty}\sum_{k^{\prime},n^{\prime}=1}^{\infty}
\tilde{M}_{kn}^{(1)}\left\{
\vphantom{\frac{kn\pi^2}{L_x^{(1)})L_y^{(1)}\gamma_{kn}^2}}
\tilde{M}_{x;k^{\prime}n^{\prime}}^{(2)}
X_{kk^{\prime}}Y_{nn^{\prime}}
\right.
\label{eq38} \\
&&~~
\left[
\frac{k^2\pi^2}{(L_x^{(1)})^2\gamma_{kn}^2}e^{-\gamma_{kn}a}
\left(1-e^{-\gamma_{kn}d_1}\right)
\left(1-e^{-\gamma_{kn}d_2}\right)\right.
\nonumber \\
&&~~
\left.
\vphantom{\frac{k^2\pi^2}{(L_x^{(1)})^2\gamma_{kn}^2}}
+{\rm Ei}(-\gamma_{kn}a)-{\rm Ei}[-\gamma_{kn}(a+d_2)]
-{\rm Ei}[-\gamma_{kn}(a+d_1)]+{\rm Ei}[-\gamma_{kn}(a+d_1+d_2)]
\right]
\nonumber \\
&&
~~~\left.
-\tilde{M}_{y;k^{\prime}n^{\prime}}^{(2)}
\tilde{X}_{kk^{\prime}}\tilde{Y}_{nn^{\prime}}
\frac{kn\pi^2}{L_x^{(1)}L_y^{(1)}\gamma_{kn}^2}
e^{-\gamma_{kn}a}
\left(1-e^{-\gamma_{kn}d_1}\right)
\left(1-e^{-\gamma_{kn}d_2}\right)
\right\}.
\nonumber
\end{eqnarray}

Similar to the case of out-of-plane magnetization,
the quantity (\ref{eq38}) is different from zero only due to
the boundary effects. The dominant contribution to Eq.~(\ref{eq38})
can be estimated using Eqs.~(\ref{eq28}) and (\ref{eq29})
and the same inequalities for $\tilde{X}_{kk^{\prime}}$ and
$\tilde{Y}_{nn^{\prime}}$ defined in Eq.~(\ref{eq37}).
As a result, we obtain that at each separation the largest
magnitude of $F_m^{\prime}$ is achieved at $\alpha=0$, i.e.,
for the parallel in-plane magnetizations. For example, when
separation increases from 223 to 400\,nm the magnitude of
$F_m^{\prime}$ decreases from $1.1\times 10^{-3}$ to
$5.3\times 10^{-4}\,\mu$N/m. Thus, in the case of in-plane
magnetization the role of magnetic interaction in dynamic
measurements of the Casimir force is even smaller than for
out-of-plane one.

Now we consider the case that there is a spontaneous magnetization in
our Ni films. This can be described by adding nonzero constant terms
$M_{00}^{(1)}$ and $M_{x,y;00}^{(2)}$ on the right-hand side
of Eq.~(\ref{eq35}). In the same way, as in Sec.~VA, for the
energy of magnetic interaction per unit area of two
parallel films we obtain (see Appendix B for details)
\begin{eqnarray}
&&
E_{sm}(a)=4M_{00}^{(1)}M_{x,00}^{(2)}\int_{a}^{a+d_2}\!\!dz_2
\left[\arctan\frac{2z_2}{\sqrt{(L_x^{(1)})^2+(L_y^{(1)})^2
+4z_2^2}}\right.
\nonumber \\
&&~~~~~
\left.
-\arctan\frac{2(z_2+d_1)}{\sqrt{(L_x^{(1)})^2+(L_y^{(1)})^2
+4(z_2+d_1)^2}}\right].
\label{eq39}
\end{eqnarray}

Using the PFA in Eq.~(\ref{eq10}), the gradient of the magnetic
force due to spontaneous magnetization takes the form
\begin{eqnarray}
&&
F_{sm}^{\prime}(a)=8\pi RM_{00}^{(1)}M_{x,00}^{(2)}
\left[\arctan\frac{2(a+d_2)}{\sqrt{(L_x^{(1)})^2+(L_y^{(1)})^2
+4(a+d_2)^2}}\right.
\nonumber \\
&&~~~
-\arctan\frac{2a}{\sqrt{(L_x^{(1)})^2+(L_y^{(1)})^2
+4a^2}}
+\arctan\frac{2(a+d_1)}{\sqrt{(L_x^{(1)})^2+(L_y^{(1)})^2
+4(a+d_1)^2}}
\nonumber \\
&&~~~
\left.
-\arctan\frac{2(a+d_1+d_2)}{\sqrt{(L_x^{(1)})^2+(L_y^{(1)})^2
+4(a+d_1+d_2)^2}}
\right].
\label{eq40}
\end{eqnarray}

Assuming that all magnetic moments are directed in one direction
(the saturation magnetization), we obtain from Eq.~(\ref{eq40})
that $|F_{sm}^{\prime}|$ increases from $1.0\times 10^{-5}$ to
$1.7\times 10^{-5}\,\mu$N/m when the separation increases from 223
to 550\,nm. This is a negligibly small effect in dynamic
measurements of the Casimir force.

\section{Comparison between experiment and theory for
${\rm\bf Ni}$ test bodies}

We have now demonstrated that possible magnetic effect due to
the domain structure of Ni films used in our experiment yields
scarcely any contribution to the measured force gradients.
Because of this the measurement results for the gradients
of the Casimir force presented in Sec.~III can be reliably
compared with the predictions of the Lifshitz theory taking
into account nonzero temperature, conductivity properties
of Ni, surface roughness and inaccuracy of the PFA, as
discussed in Sec.~IV.

In Ref.\cite{48} we have used the traditional method of the
comparison between experiment and theory when the measurement
data are presented as crosses whose horizontal arms are
equal to $2\Delta a$ and the vertical arms are equal to
$2\Delta^{\! t}F^{\prime}(a)$. Here we use another method
of comparison\cite{9,10,43,52} based on consideration of the
confidence interval for the random quantity
$F_{R}^{\prime}(a_i)-\bar{F}^{\prime}(a_i)$ equal to the
difference between theoretical and mean experimental force
gradients at the experimental separations $a_i$.
This method is  advantageous because it allows to make the
quantitative conclusions not only about the rejection of
any theoretical approach, but about the measure of agreement
between experiment and theory as well.

To calculate the confidence interval for the difference
between theoretical and mean experimental force
gradients, one needs to have the total errors of both
quantities.  The total experimental error
$\Delta^{\! t}F^{\prime}$ is already determined in Sec.~III
(see Fig.~\ref{fg3}). The crucial contribution to the
theoretical error is given by the errors in optical data
of Ni determined by the number of significant figures in
the tables.\cite{53}  The errors in the optical data lead
to the theoretical error $\Delta^{\! \rm opt}F_{R}^{\prime}$
equal to approximately 0.5\% of $F_{R}^{\prime}$ (it is
shown by the long-dashed line in Fig.~\ref{fg7} as a function
of separation). There is, however, one more source of
error\cite{86} when the theoretical value of the force
gradient is calculated not over some separation interval
but at the experimental separations $a_i$.
The point is that each experimental separation is determined
up to an error $\Delta a$ and this leads to a respective
error in the calculated force gradients\cite{86}
\begin{equation}
\Delta^{\!\rm sep}F_{R}^{\prime}(a_i)
\approx 4\frac{\Delta a}{a_i}F_{R}^{\prime}(a_i).
\label{eq41}
\end{equation}
\noindent
In Fig.~\ref{fg7} the theoretical error
$\Delta^{\!\rm sep}F_{R}^{\prime}$ as a function of separation
is shown by the short-dashed line.
In the same figure the solid line shows the total theoretical
error $\Delta^{\! t}F_{R}^{\prime}$ determined at a 67\%
confidence level which was combined in quadrature from the
theoretical errors $\Delta^{\!\rm opt}F_{R}^{\prime}$  and
$\Delta^{\!\rm sep}F_{R}^{\prime}$.
The total theoretical error varies from 0.99 to $0.03\,\mu$N/m
when the separation increases from 223 to 500\,nm.
As can be seen from the comparison with Fig.~\ref{fg3},
at all separations the total experimental error is in exceess
of the total theoretical error. As a result, the confidence
interval for the quantityv
$F_{R}^{\prime}(a_i)-\bar{F}^{\prime}(a_i)$  determined at a
67\% confidence level is given by
$[-\Xi_{F^{\prime}}^{0.67}(a),\Xi_{F^{\prime}}^{0.67}(a)]$,
where
\begin{equation}
\Xi_{F^{\prime}}^{0.67}(a)=\left\{
[\Delta^{\! t}F^{\prime}(a)]^2+
[\Delta^{\! t}F_{R}^{\prime}(a)]^2
\right\}^{1/2}.
\label{eq42}
\end{equation}

In Fig.~\ref{fg8}(a) we show the quantities
$\pm\Xi_{F^{\prime}}^{0.67}(a)$ as functions of separation by
the solid lines. In doing so the confidence interval at each fixed
$a$ is the vertical segment between $-\Xi_{F^{\prime}}^{0.67}(a)$
and $\Xi_{F^{\prime}}^{0.67}(a)$. It has the meaning that if the
theory is consistent with the data then at least 67\% data points
within each separation subinterval must belong to this confidence
interval. To verify which of the two theoretical approaches used
in Sec.~IV is consistent with the data, in Fig.~\ref{fg8}(a,b)
we plot the differences
$F_{R}^{\prime}(a_i)-\bar{F}^{\prime}(a_i)$  as black and gray
dots, where the plasma model and the Drude model approaches,
respectively, were used to compute the quantity
$F_{R}^{\prime}(a_i)$
(see the lower and upper lines in Fig.~\ref{fg5}).
As can be seen in Fig.~\ref{fg8}(a), not only 67\% but all
black dots belong to the confidence intervals within the entire
separation region from 223 to 550\,nm. This means that the data
are consistent with the Lifshitz theory combined with the plasma
model approach with a large safety margin.
As to the gray dots, most of them are outside the confidence
intervals over the separation region from 223 to 420\,nm.
Thus, the Drude model approach to the Casimir force is excluded
by the data within this separation region at a 67\% confidence
level.

As is seen in Fig.~\ref{fg8}(a), even if the confidence intervals
were widened to reach a 95\% confidence level, the differences
$F_{R,D}^{\prime}(a_i)-\bar{F}^{\prime}(a_i)$ computed using the
Drude model approach would still remain outside those intervals
within some range of separations. To make this observation
quantitative, we calculate the half-width of the confidence
interval $\Xi_{F^{\prime}}^{0.95}(a)$ from the equation
\begin{equation}
\frac{\Xi_{F^{\prime}}^{0.95}(a)}{\Xi_{F^{\prime}}^{0.67}(a)}=
\frac{t_{(1+0.95)/2}(32)}{t_{(1+0.67)/2}(32)}\approx 2.
\label{eq43}
\end{equation}
\noindent
In Fig.~\ref{fg8}(b) we plot
the borders of the 95\% confidence intervals
$\pm\Xi_{F^{\prime}}^{0.95}(a)$ by the solid lines and reproduce
the black and gray dots from Fig.~\ref{fg8}(a). As can be seen in
Fig.~\ref{fg8}(b), the most of gray dots are still outside the
widened confidence intervals within the separation interval from
223 to 345\,nm. This allows one to conclude that at these separations
the Drude model approach is excluded by the data at a higher,
95\%, confidence level.

To give a better understanding of the character of agreement
(disagreement) between the nonaveraged data and two
theoretical approaches, in Fig.~\ref{fg9} we provide a
histogram plotted at $a=251\,$nm. Here, $f$ is the fraction of
33 data points having $F^{\prime}$ in the bin shown by the
respective vertical lines. The data are consistent with the
Gaussian distribution with the standard deviation
$\sigma_{F^{\prime}}=0.92\,\mu$N/m and the mean
gradient $\bar{F}^{\prime}=74.17\,\mu$N/m
shown by the dashed line. The black and gray vertical lines in
Fig.~\ref{fg9} show the theoretical predictions of the plasma
model, ${F}_{R,p}^{\prime}=74.19\,\mu$N/m, and the Drude model,
${F}_{R,D}^{\prime}=77.46\,\mu$N/m, approaches, respectively.
Note that in Ref.~\cite{48} there is a typo in the value of
separation (250\,nm instead of 251\,nm in the inset to Fig.~2).
It is seen that the plasma model approach is
in a very good agreement with the measurement result as
\begin{equation}
{F}_{R,p}^{\prime}-\bar{F}^{\prime}=0.02\,\mu\mbox{N/m}
<\frac{1}{52}\Delta^{\! t}{F}^{\prime}.
\label{eq44}
\end{equation}
\noindent
At the same time the theoretical prediction of the Drude model
approach is excluded at high confidence as
\begin{equation}
{F}_{R,D}^{\prime}-\bar{F}^{\prime}=3.29\,\mu\mbox{N/m}
>3\Delta^{\! t}{F}^{\prime}.
\label{eq45}
\end{equation}

In the end of this section we emphasize that although
Figs.~\ref{fg8}(a) and (b) allow the exclusion of the Drude model
approach in a quantitative way (at 67\% and 95\% confidence
levels within respective separation regions), they cannot be
considered as a confirmation of the plasma model approach
at either 67\% or 95\% confidence level. The situation here is
just the opposite: the higher is the confidence level at which
the Drude model is excluded [for example, 95\% in
Fig.~\ref{fg8}(b)],
the easier is for the plasma model approach to accomodate all
the points for the gradient differences within the widened
confidence interval. In fact, to obtain the quantitative
description for the measure of agreement between some
experimentally consistent theoretical approach and the data
one should make the confidence interval as narrow as possible
and determine the respective low confidence level at which
this approach is excluded by the data. Then one can conclude
that the theoretical approach under consideration is confirmed
by the data at a complementary to 100\% high confidence.

To illustrate the above, let us consider the confidence interval
 $[-\Xi_{F^{\prime}}^{0.1}(a),\Xi_{F^{\prime}}^{0.1}(a)]$
 defined at a 10\% confidence.
This can be found from the equality
\begin{equation}
\frac{\Xi_{F^{\prime}}^{0.1}(a)}{\Xi_{F^{\prime}}^{0.67}(a)}=
\frac{t_{(1+0.1)/2}(32)}{t_{(1+0.67)/2}(32)}\approx 0.13.
\label{eq46}
\end{equation}
\noindent
In Fig.~\ref{fg10} the borders of the 10\% confidence intervals
are plotted as the two solid lines and the black dots show the
same differences
${F}_{R,p}^{\prime}(a_i)-\bar{F}^{\prime}(a_i)$ as are shown by
the black dots in Fig.~\ref{fg8}.
As can be seen in Fig.~\ref{fg10}, in spite of rather narrow
10\% confidence intervals, much more than 10\% of all dots
within any separation subinterval belong to them.
This means that the plasma model approach is not excluded by the
data even at a 10\% confidence level or, equivalently, that
this approach is confirmed by the data at more than 90\%
confidence level.

As mentioned in Secs.~I and IV, it remains unclear why the
experimental data are in agreement with theory disregarding
really existing relaxation properties of conduction electrons
and exclude the theory taking these properties into account.
In the next section we compare the above experimental results
for two Ni films with respective measurements in configurations
containing one or two Au test bodies. We show that the unique
feature of two Ni test bodies shown in Fig.~\ref{fg5}
$({F}_{R,D}^{\prime}>{F}_{R,p}^{\prime})$ leads to important
conclusions with respect to the role of possible background
effects in measurements of the Casimir force, such as patch
potentials.\cite{87}

\section{Comparison with experiments involving nonmagnetic
metals}
Here we compare the experimental results and the measure of their
agreement with theory for two Ni test bodies with the results of
previous measurements using the same setup. One of them was
performed\cite{39,88} with an Au-coated plate and an Au-coated
sphere, and the other\cite{32} with a Ni-coated plate and
an Au-coated sphere.

We begin from the experiment\cite{39,88} using an Au-coated plate
and an Au-coated sphere of $R=41.3\,\mu$m radius.
First, we present the results of this experiment in terms of
the differences between the theoretical force gradients computed
using either the Drude model or the plasma model approaches and
mean measured gradients of the Casimir force.
These differences are shown in Fig.~\ref{fg11}(a,b) by the gray
and black dots, respectively.
Note that for Au computations using the Drude model approach
have been made\cite{39,88} with the tabulated optical
data\cite{53}
extrapolated to zero frequency by the Drude model (\ref{eq13})
with the parameters $\omega_p=9.0\,$eV and $\gamma=0.035\,$eV.
Recently it was shown\cite{89} that $\varepsilon(i\xi_l)$
obtained in this way is in excellent agreement with the
dielectric permittivity obtained by means of the weighted
Kramers-Kronig relations from the tabulated optical data\cite{53}
with no extrapolation. Furthermore, ellipsometry measurements
of the optical properties of Au films were found\cite{90}
in good agreement with the results of Ref.\cite{53}.
The alternative optical data for Au contained in the literature,
which can significantly deviate from the tabulated data,\cite{53}
were shown\cite{9,10} to lead to much larger deviations between the
predictions of the Drude model approach and measurements of the
Casimir force than the data of Ref.\cite{53}.

The solid lines in Fig.~\ref{fg11} indicate the borders of the
confidence intervals determined at (a) 67\% confidence and
(b) 95\% confidence level. They are found by using the total
experimental and theoretical errors in the experiment of
Refs.\cite{39,88}, as discussed in Sec.~VI of the present
paper. As can be seen in Fig.~\ref{fg11}, the plasma model
approach is consistent with the data over the entire
separation region. As to the Drude model approach, it is
excluded by the data at a 67\% confidence level over the
separation region from 235 to 420\,nm
[see Fig.~\ref{fg11}(a)] and
at a 95\% confidence level over the
separation region from 235 to 330\,nm
[see Fig.~\ref{fg11}(b)].

{}From the comparison of Figs.~\ref{fg8} and \ref{fg11} one
can observe an important difference between the cases of
Ni-Ni and Au-Au test bodies. Note that for Ni-Ni
test bodies $F_{R,D}^{\prime}-\bar{F}^{\prime}>0$
(see Fig.~\ref{fg8}), i.e..,
$F_{R,D}^{\prime}>{F}_{R,p}^{\prime}$,
in contrast to the case of Au-Au test bodies where
$F_{R,D}^{\prime}-\bar{F}^{\prime}<0$
(see Fig.~\ref{fg11}) and
$F_{R,D}^{\prime}<{F}_{R,p}^{\prime}$.
This difference sheds light on the possible role and size
of electrostatic patches in measurements of the Casimir force.
It was hypothesized\cite{91} that an additional attractive
force due to the effect of large patches might bring the
experimental data for the two Au test bodies in agreement with
the predictions of the Drude model approach.
{}From Fig.~\ref{fg11} it is seen that the attractive force
with a magnitude equal to the difference between two sets
of dots would really bring the gray dots in agreement and
the black dots in disagreement with the data.
It is not logical, however, to assume that the patch
effect plays this role for Au but does not play the same
role for Ni.
{}From Fig.~\ref{fg8} it follows that any additional attractive
force would only increase the disagreement of the Drude model
approach with the data leading also to a disagreement of the
plasma model approach with the same data.
This is in favor of the statement that surface patches lead
to only a negligibly small effect in measurements of the
Casimir interaction by means of AFM and micromachined
oscillator\cite{9,10,43} in qualitative agreement to the
model of patches proposed in Ref.\cite{87}.
This conlusion was recently confirmed\cite{92} by means of
Kelvin probe microscopy.

Futher confirmation for a negligibly small role of the effect
of electrostatic patches in measurements of the Casimir
interaction by means of an AFM comes from the experiment\cite{32}
with a Ni-coated plate and an Au-coated sphere of
$R=64.1\,\mu$m radius. In this configuration the predictions of both
theoretical approaches to the Casimir force are almost coincident
over the experimental separations range. To see this in
Fig.~\ref{fg12}(a) we show the quantity $F_{\rm diff}^{\prime}(a)/R$
[see Eq.~(\ref{eq16})] by the three solid lines from top to
bottom for the experiments on measuring the gradient of the
Casimir force between Ni-Ni (this work), Ni-Au (Ref.~\cite{32})
and Au-Au test bodies,\cite{39,88} respectively (in each case
the respective value of the sphere radius is used to make the
presented results comparable). To gain a better understanding
of distinctions between the two theoretical approaches, in
Fig.~\ref{fg12}(b) we also show the quantity
$F_{\rm diff}^{\prime}(a)/F_{R,p}^{\prime}(a)$ in percent
for the three experiments in the same succession as  in
Fig.~\ref{fg12}(a). As is seen  in Fig.~\ref{fg12}(a,b),
for Ni-Au test bodies (the lines sandwiched between the top
and bottom ones) the quantity $F_{\rm diff}^{\prime}/R$ and
$F_{\rm diff}^{\prime}/F_{R,p}^{\prime}$ cannot be
distinguished from zero in the limits of experimental errors.
However, for all the three experiments, including that with
Ni-Au test bodies, the measurement results are consistent with
theoretical predictions using the plasma model approach.
This is seen in Fig.~\ref{fg13}(a,b) where, to make the results
of the different experiments comparable, the quantity
$F_{R,p}^{\prime}/R$ is shown by the solid dark bands and the
crosses represent measurement data
with their total experimental errors
normalized by the radii (the bands having
thicknesses equal to twice the theoretical error are again plotted
from top to bottom for experiments with Au-Au, Ni-Au, and Ni-Ni
test bodies, respectively). Remembering that for Ni-Au test
bodies two alternative theoretical approaches lead to almost
coincident predictions, an introduction of some detectable
additional force originating, for instance, from patch potentials
would inevitably make both approaches inconsistent with the
experimental data.

In Fig.~\ref{fg14} we demonstrate that it is impossible to
simultaneously reconcile the Drude model approach with the data
of two experiments using Au-Au and Ni-Ni test bodies at the
expense of any unaccounted hypothetical background effect
leading to either attractive or repulsive force.
In this figure the upper and lower bands show the theoretical
results obtained using the Drude model approach for the
quantity $F_{R,D}^{\prime}/R$ for Au-Au and Ni-Ni test bodies,
respectively. As can be seen in Fig.~\ref{fg14}, there is an
evident inconsistency between the data of both experiments and
theoretical predictions of the Drude model approach.
The important point is that to remedy the problem one would
need to introduce some hypothetical attractive force for the
experiment with Au-Au test bodies (the upper band and set of
crosses) and a hypothetical repulsive force for the
experiment with Ni-Ni test bodies (the lower band and set of
crosses). Thus, not only an electrostatic attraction due to
patch potentials, but any unaccounted hypothetical
interaction preserving its sign (i.e., being either
attractive or repulsive) is incapable to reconcile the
predictions of the Drude model approach with the data.
Keeping in mind that in Sec.~V we have carefully examined
possible contributions of magnetic interactions due to the
domain structure of Ni films and found it negligibly small,
any alternative interpretation of our measurement results
faces severe difficulties.

\section{Conclusions and discussion}

In the foregoing we have presented complete calibration and
measurement data of the experiment on measuring the
gradient of the Casimir force between a Ni-coated plate and
a Ni-coated sphere by means of dymanic AFM operated in the
frequency shift technique. This is the pioneering
experiment which measured the influence of magnetic
properties of the boundary metals on the Casimir interaction
predicted theoretically more than 40 years ago.
Taking into account that the magnitudes of the force gradients
under consideration are about or less than $100\,\mu$N/m
and the magnetic properties contribute up to 5\% of this
quantity, it becomes clear that such experiments call
for extreme care to the vacuum system, surface preparation,
calibration procedures, background effects, error analysis
and comparison between experiment and theory.
In this paper we have presented exhaustive information
on all the above subjects which has not been already
elucidated in the papers devoted to previous experiments
using the same setup with Au-Au and Au-Ni test
bodies,\cite{32,39,88} and with Ni-Ni test bodies published
only in Letter form.\cite{48}

After a brief description of some details of the setup which
were not described in the literature so far, we have presented
the results of the electrostatic calibrations which allow
precise determination of the calibration constant, closest
absolute separation and residual potential difference.
All the details of error analysis, including the random,
systematic errors and their combination into the total
error, were provided.
Both individual measured gradients of the Casimir force and
their mean values were presented. Computations of the
gradients of the Casimir force in the sphere-plane geometry were
performed using the Lifshitz theory at nonzero temperature
taking into account the recently calculated correction terms
to the PFA and the surface roughness. The conductivity
properties of Ni were described in succession using the Drude
and the plasma model approaches to the Casimir force presented
in the literature and the obtained results were compared
between themselves and with the total experimental errors.

We have investigated possible magnetic interaction
between the test bodies in our experimental configuration
arising due to the domain structure of Ni films. Both cases,
 out-of-plane and in-plane magnetizations, have been studied
extensively (the former has been only briefly
considered with respect to
measurements of the Casimir force\cite{48} and the latter
was not previously investigated).
Although extreme care has been taken in order to avoid
spontaneous magnetization of the Ni films used, the case of
the fully magnetized films was also considered. It was shown
that in all cases the contribution of magnetic interaction to
the measured force gradient is by several orders of magnitude
smaller than the total experimental error. This allowed
a reliable comparison of the measured gradients of the
Casimir force with theoretical predictions.

The comparison of the experimental results with theory was
based on a more rigorous method different from that used in
Refs.\cite{32,39,48,88}. This method is based on the
consideration of the random quantity equal to the difference
between theoretical and mean experimental force gradients.
Both 67\% and 95\% confidence intervals for this quantity were
found. The preference of the comparison method under discussion
is that it not only allows one to exclude some theoretical
approach as inconsistent with the data at a given confidence
level, but also permits to quantitatively determine at what
confidence level a theoretical approach is confirmed by the
data. On this basis we have concluded that the Drude model
approach to the Casimir force is excluded by our measurements
with two Ni surfaces at a 95\% confidence level, whereas the
plasma model approach is confirmed  by the data at higher than
90\% confidence level. In this work we have
investigated in detail
the striking property of the Casimir interaction
between two magnetic test bodies, i.e., that the force
gradients calculated using the Drude model approach are
significantly larger than the measured mean force gradients.
This is just the opposite of the case of two nonmagnetic
(Au) test bodies where the theoretical force gradients,
calculated using the Drude model approach are
significantly smaller than the measured mean force gradients.
By comparing the measurement results of the three experiments
with Au-Au, Ni-Au, and Ni-Ni test bodies taking the above
property into account, we have arrived at the conclusion
of major importance that no hypothetical unaccounted
background force (either attractive or repulsive) could
bring the measurement data into agreement with theoretical
predictions of the Drude model approach (the attractive
force arising due to electrostatic patches is only one
example of possible interactions).
This means that an exclusion of the
Drude model approach by the data assumes a greater significance
which awaits for its fundamental
explanation.\cite{72}

To conclude we would like to stress that the experiment on
measuring the gradient of the Casimir force between two
Ni surfaces has brought confirmation to the prediction
of the Lifshitz theory that magnetic properties of boundary
surfaces influence the Casimir force.
According to our measurement results, the quantitative
description of the Casimir interaction between both
magnetic and nonmagnetic metals is given by the plasma
model approach. At this point it is pertinent to note that in the
configuration of a ferromagnetic dielectric interacting
with a nonmagnetic metal described by the plasma model the
Lifshitz theory predicts the Casimir repulsion through a
vacuum gap.\cite{24,25,26}
This makes possible realization of the Casimir repulsion on
microscales in the near future for subsequent applications to the
problems of lubrication and friction in nanodevices.

\section*{Acknowledgments}
The authors are grateful to G.~Bimonte for providing the
numerical values of a correction beyond the PFA at
room temperature
from Fig.~2 of Ref.\cite{76} and to
L.~P.~Teo for providing the
numerical values of the same correction
at $T=0$
from Fig.~9 of Ref.\cite{77}.
This work was supported by the DOE Grant
No.~DEF010204ER46131 (equipment, G.L.K., V.M.M., U.M.).
\section*{APPENDIX A}
\renewcommand{\theequation}{A\arabic{equation}}
\setcounter{equation}{0}

In this Appendix we derive some mathematical results used in
Sec.~VA to calculate the magnitude of magnetic interactions
in our experimental setup for out-of-plane magnetized Ni
films.

{}From Eq.~(\ref{eq17}) we can find the $z$-component of the
magnetic field created by the periodically extended first Ni
film at the points of the second Ni film
\begin{equation}
H_z(x_2,y_2,z_2)=\int_{-\infty}^{\infty}\!\!\!dx_1
\int_{-\infty}^{\infty}\!\!\!dy_1\int_{-d_1}^{0}\!\!\!dz_1
\left[\frac{3(z_2-z_1)^2}{|\mbox{\boldmath$r$}|^5}-
\frac{1}{|\mbox{\boldmath$r$}|^3}\right]M_z^{(1)}(x_1,y_1),
\label{A1}
\end{equation}
\noindent
where the radius-vector {\boldmath$r$} is defined in
Eq.~(\ref{eq19}) and the magnetization is specified in
Eq.~(\ref{eq21}).
Calculating the integral with respect to $z_1$ in
Eq.~(\ref{A1}), we obtain
\begin{eqnarray}
&&
H_z(x_2,y_2,z_2)=\int_{-\infty}^{\infty}\!\!\!dx_1
\int_{-\infty}^{\infty}\!\!\!dy_1\left\{
\frac{z_2}{\left[z_2^2+(x_2-x_1)^2+(y_2-y_1)^2\right]^{3/2}}
\right.
\label{A2} \\
&&
~~~
\left.-
\frac{z_2+d_1}{\left[(z_2+d_1)^2+(x_2-x_1)^2+(y_2-y_1)^2\right]^{3/2}}
\right\}M_z^{(1)}(x_1,y_1).
\nonumber
\end{eqnarray}
\noindent
Now we assume that there is no spontaneous magnetization and
substitute the Fourier series (\ref{eq22}) in Eq.~(\ref{A2}).
After introducing the new variables $u=x_1-x_2$ and
$v=y_1-y_2$, transforming the sinus functions and equating to
zero the integrals of odd functions, Eq.~(\ref{A2}) can be
brought to the form
\begin{eqnarray}
&&
H_z(x_2,y_2,z_2)=\sum_{k,n=1}^{\infty}
M_{kn}^{(1)}\sin\frac{k\pi x_2}{L_x^{(1)}}
\sin\frac{n\pi y_2}{L_y^{(1)}}
\nonumber \\
&&
~~~~~~\times
\left[z_2\Phi_{kn}(z_2)-(z_2+d_1)\Phi_{kn}(z_2+d_1)\right].
\label{A3}
\end{eqnarray}
\noindent
Here we have introduced the notation
\begin{equation}
\Phi_{kn}(z)=\int_{-\infty}^{\infty}\!\!\!du
\int_{-\infty}^{\infty}\!\!\!dv
\frac{\cos\frac{k\pi u}{L_x^{(1)}}
\cos\frac{n\pi v}{L_y^{(1)}}}{(u^2+v^2+z^2)^{3/2}}.
\label{A4}
\end{equation}

The double integral in Eq.~(\ref{A4}) can be evaluated
explicitly. For this purpose we set
$\pi k/L_x^{(1)}=a_k$, $\pi n/L_y^{(1)}=b_k$ and
calculate the derivative\cite{85}
\begin{eqnarray}
\frac{d\Phi_{kn}(z)}{d b_n}&=&-2
\int_{-\infty}^{\infty}\!\!\!du\cos(a_ku)
\int_{0}^{\infty}\!\!\!dv
\frac{v\sin(b_nv)}{(u^2+z^2+v^2)^{3/2}}
\nonumber \\
&=&-4b_n
\int_{-\infty}^{\infty}\!\!\!du\cos(a_ku)
K_0(b_n\sqrt{z^2+u^2})
\nonumber \\
&=&-2\pi
\frac{b_n}{\gamma_{kn}}
e^{-\gamma_{kn}z},
\label{A5}
\end{eqnarray}
\noindent
where $K_0(t)=({\pi i}/{2})H_0^{(1)}(it)$ is the Bessel function
of imaginary argument and $\gamma_{kn}$ is defined in
Eq.~(\ref{eq25}).
Then by the integration of Eq.~(\ref{A5}) with
respect to $b_n$ one finds
\begin{equation}
\Phi_{kn}(z)=\frac{2\pi}{z}e^{-\gamma_{kn}z}+
G(a_k,z),
\label{A6}
\end{equation}
\noindent
where $G(a_k,z)$ is the integration constant.
The value of this constant can be found by considering the
quantity (\ref{A4}) with $n=b_n=0$
\begin{eqnarray}
\Phi_{k0}(z)&=&4
\int_{0}^{\infty}\!\!\!du\cos(a_ku)
\int_{0}^{\infty}\!\!
\frac{dv}{(u^2+z^2+v^2)^{3/2}}
\nonumber \\
&=&4
\int_{0}^{\infty}\!\!\!du\frac{\cos(a_ku)}{z^2+u^2}=
\frac{2\pi}{z}e^{-za_k}.
\label{A7}
\end{eqnarray}
\noindent
Comparing this with  Eq.~(\ref{A6}), we can conclude that
$G(a_k,z)=0$. Thus, from (\ref{A6}) one arrives  at the
final expression (\ref{eq25}) for the function $\Phi_{kn}(z)$.

The energy of magnetic interaction between parallel plates can
be now obtained from Eq.~(\ref{eq20})
\begin{equation}
E_m(a)=-\int_{0}^{L_x^{(1)}}\!\!\!dx_2
\int_{0}^{L_y^{(1)}}\!\!\!dy_2
\int_{a}^{a+d_2}\!\!\!dz_2M_z^{(2)}(x_2,y_2)
H_z(x_2,y_2,z_2).
\label{A8}
\end{equation}
\noindent
Substituting here with Eq.~(\ref{eq22}) for the magnetization of the
second film, Eq.~(\ref{A3}) for the magnetic field and using
notations (\ref{eq24}), one arrives at the expression in
Eq.~(\ref{eq23}).

If the spontaneous magnetizaion is present, Eq.~(\ref{A2})
for the respective magnetic field created by the first film,
should be rewritten in the form
\begin{eqnarray}
&&
H_z(x_2,y_2,z_2)=M_{00}^{(1)}
\int_{-L_x^{(1)}/2}^{L_x^{(1)}/2}\!\!\!dx_1
\int_{-L_y^{(1)}/2}^{L_y^{(1)}/2}\!\!\!dy_1
\left\{
\frac{z_2}{\left[z_2^2+(x_2-x_1)^2+(y_2-y_1)^2\right]^{3/2}}
\right.
\nonumber \\
&&
~~~
\left.-
\frac{z_2+d_1}{\left[(z_2+d_1)^2+(x_2-x_1)^2+(y_2-y_1)^2\right]^{3/2}}
\right\}.
\label{A9}
\end{eqnarray}
\noindent
Now  we take into account that the second film is situated above
the center of a large plate, i.e., $x_2<< L_x^{(1)}$ and
$y_2<< L_y^{(1)}$. Thus, with sufficient precision one can put
$x_2\approx y_2\approx 0$. Replacing the first film with a disc
of $L_x^{(1)}/2=0.5\,$cm radius, we obtain the following
estimate
\begin{equation}
H_z(z_2)\approx 4\pi M_{00}^{(1)}\left[
\frac{z_2+d_1}{\sqrt{(L_x^{(1)})^2+4(z_2+d_1)^2}}
-\frac{z_2}{\sqrt{(L_x^{(1)})^2+4z_2^2}}\right].
\label{A10}
\end{equation}
\noindent
Then, calculating the magnetic energy arising per unit film area
due to the spontaneous magnetization
\begin{equation}
E_{sm}(a)=-M_{00}^{(2)}\int_{a}^{a+d_2}\!\!\!dz_2H_z(z_2),
\label{A11}
\end{equation}
\noindent
we arrive at Eq.~(\ref{eq30}).

\section*{APPENDIX B}
\renewcommand{\theequation}{B\arabic{equation}}
\setcounter{equation}{0}

Here we derive the mathematical expressions used in Sec.~VB to
calculate the gradient of magnetic force for the case of
in-plane magnetization of Ni films.

We begin from calculation of the $x$- and $y$-components of
magnetic field created by the periodically continued first
Ni film at the points of parallel to it second Ni film.
From Eq.~(\ref{eq17}) for the in-plane magnetization one obtains
\begin{eqnarray}
&&
H_x(x_2,y_2,z_2)=\int_{-\infty}^{\infty}\!\!\!dx_1
\int_{-\infty}^{\infty}\!\!\!dy_1\int_{-d_1}^{0}\!\!\!dz_1
\left[\frac{3(x_2-x_1)^2}{|\mbox{\boldmath$r$}|^5}-
\frac{1}{|\mbox{\boldmath$r$}|^3}\right]M_x^{(1)}(x_1,y_1),
\nonumber \\
&&
H_y(x_2,y_2,z_2)=\int_{-\infty}^{\infty}\!\!\!dx_1
\int_{-\infty}^{\infty}\!\!\!dy_1\int_{-d_1}^{0}\!\!\!dz_1
\frac{3(x_2-x_1)(y_2-y_1)}{|\mbox{\boldmath$r$}|^5}
M_x^{(1)}(x_1,y_1),
\label{B1}
\end{eqnarray}
\noindent
where the magnetization is presented in Eq.~(\ref{eq35}).
Let us calculate the component $H_y$ first. For this purpose
we use the identity
\begin{equation}
\frac{3(x_2-x_1)}{|\mbox{\boldmath$r$}|^5}=
-\frac{\partial}{\partial x_2}
\frac{1}{\left[(x_2-x_1)^2+(y_2-y_1)^2+
(z_2-z_1)^2\right]^{3/2}}.
\label{B2}
\end{equation}
\noindent
Substituting Eqs.~(\ref{B2}) and (\ref{eq35}) in Eq.~(\ref{B1})
and using the variables $u$ and $v$ introduced in Appendix A,
we find
\begin{eqnarray}
&&
H_y(x_2,y_2,z_2)=-\sum_{k,n=1}^{\infty}\tilde{M}_{kn}^{(1)}
\frac{\partial}{\partial x_2}
\int_{-\infty}^{\infty}\!\!\!du
\int_{-\infty}^{\infty}\!\!\!dv\int_{-d_1}^{0}\!\!\!dz_1
\nonumber \\
&&~~~
\times\frac{v\sin[a_k(u+x_2)]\sin[b_n(v+y_2)]}{[u^2+v^2+
(z_2-z_1)^2]^{3/2}},
\label{B3}
\end{eqnarray}
\noindent
where $a_k$ and $b_n$ are defined in Appendix A below
Eq.~(\ref{A4}). Now we transform the sinus functions, set
equal to zero the integrals of odd functions, and
calculate the derivative with respect to $x_2$.
The result is
\begin{eqnarray}
&&
H_y(x_2,y_2,z_2)=-\sum_{k,n=1}^{\infty}\tilde{M}_{kn}^{(1)}
a_k\cos(a_kx_2)\cos(b_ny_2)
\nonumber \\
&&~~~
\times
\int_{-d_1}^{0}\!\!\!dz_1
\int_{-\infty}^{\infty}\!\!\!du
\int_{-\infty}^{\infty}\!\!\!dv
\frac{v\cos(a_ku)\sin(b_nv)}{[u^2+v^2+
(z_2-z_1)^2]^{3/2}}.
\label{B4}
\end{eqnarray}
\noindent
Using the differentiation with respect to $b_n$ and the
notation (\ref{A4}), Eq.~(\ref{B4}) can be identically presented
in the form
\begin{eqnarray}
&&
H_y(x_2,y_2,z_2)=\sum_{k,n=1}^{\infty}\tilde{M}_{kn}^{(1)}
a_k\cos(a_kx_2)\cos(b_ny_2)
\nonumber \\
&&~~~
\times
\int_{-d_1}^{0}\!\!\!dz_1
\frac{\partial}{\partial b_n}\Phi_{kn}(z_2-z_1).
\label{B5}
\end{eqnarray}
\noindent
Substituting here Eq.~(\ref{A5}) one obtains after some
transformations
\begin{eqnarray}
&&
H_y(x_2,y_2,z_2)=-2\pi\sum_{k,n=1}^{\infty}\tilde{M}_{kn}^{(1)}
\frac{a_kb_n}{\gamma_{kn}}\cos(a_kx_2)\cos(b_ny_2)
\nonumber \\
&&~~~
\times
e^{-\gamma_{kn}z_2}
\int_{-d_1}^{0}\!\!\!dz_1
e^{\gamma_{kn}z_1}
\label{B6}
\end{eqnarray}
\noindent
leading to the final expression
\begin{eqnarray}
&&
H_y(x_2,y_2,z_2)=-2\pi\sum_{k,n=1}^{\infty}\tilde{M}_{kn}^{(1)}
\frac{a_kb_n}{\gamma_{kn}}
e^{-\gamma_{kn}z_2}(1-e^{-\gamma_{kn}d_1})
\nonumber \\
&&~~~
\times
\cos(a_kx_2)\cos(b_ny_2).
\label{B7}
\end{eqnarray}

In a similar way the component $H_x$ from Eq.~(\ref{B1}) can
be written in the form
\begin{eqnarray}
&&
H_x(x_2,y_2,z_2)=-\sum_{k,n=1}^{\infty}\tilde{M}_{kn}^{(1)}
\int_{-\infty}^{\infty}\!\!\!dx_1
\int_{-\infty}^{\infty}\!\!\!dy_1\int_{-d_1}^{0}\!\!\!dz_1
\nonumber \\
&&~~~
\times\left\{(x_2-x_1)
\frac{\partial}{\partial x_2}
\frac{\sin(a_kx_1)\sin(b_ny_1)}{\left[(x_2-x_1)^2+(y_2-y_1)^2+
(z_2-z_1)^2\right]^{3/2}}\right.
\nonumber \\
&&~~~~~~
\left. +
\frac{\sin(a_kx_1)\sin(b_ny_1)}{\left[(x_2-x_1)^2+(y_2-y_1)^2+
(z_2-z_1)^2\right]^{3/2}}\right\}.
\label{B8}
\end{eqnarray}
\noindent
Then we again introduce the new variables $u$ and $v$, set to
zero the integrals of odd functions, calculate the derivative
with respect to $x_2$ and introduce the derivative with respect
to $a_k$ in order to use Eqs.~(\ref{A4}) and (\ref{A5}).
These allow the following representation of Eq.~(\ref{B8}):
\begin{eqnarray}
&&
H_x(x_2,y_2,z_2)=-2\pi\sum_{k,n=1}^{\infty}\tilde{M}_{kn}^{(1)}
\sin(a_kx_2)\sin(b_ny_2)
\nonumber \\
&&~
\times\left[-\frac{a_k^2}{\gamma_{kn}}
\int_{-d_1}^{0}\!\!\!dz_1e^{-\gamma_{kn}(z_2-z_1)}+
\int_{-d_1}^{0}\!\frac{dz_1}{z_2-z_1}e^{-\gamma_{kn}(z_2-z_1)}
\right].
\label{B9}
\end{eqnarray}
\noindent
After integration and identical transformations one finally
obtains
\begin{eqnarray}
&&
H_x(x_2,y_2,z_2)=2\pi\sum_{k,n=1}^{\infty}\tilde{M}_{kn}^{(1)}
\sin(a_kx_2)\sin(b_ny_2)
\nonumber \\
&&~
\times\left\{\frac{a_k^2}{\gamma_{kn}^2}
e^{-\gamma_{kn}z_2}(1-e^{-\gamma_{kn}d_1})+
{\rm Ei}(-\gamma_{kn}z_2)-{\rm Ei}[-\gamma_{kn}(z_2+d_1)]\right\}.
\label{B10}
\end{eqnarray}

The magnetic energy between two parallel plates with in-plane
magnetization is obtained from Eq.~(\ref{eq20})
\begin{equation}
E_m(a)=-\int_{0}^{L_x^{(1)}}\!\!\!dx_2
\int_{0}^{L_y^{(1)}}\!\!\!dy_2
\int_{a}^{a+d_2}\!\!\!dz_2
\left(M_x^{(2)}H_x+M_y^{(2)}H_y\right).
\label{B11}
\end{equation}
\noindent
Substituting Eqs.~(\ref{eq35}), (\ref{B7}) and  (\ref{B10}) in
Eq.~(\ref{B11}), one arrives at Eq.~(\ref{eq36}).

In the end we consider the case when the spontaneous
magnetization is not equal to zero. We can again assume that
the second film is situated above the center of the first and
put $x_2\approx y_2\approx 0$. From symmetry considerations it
also follows that $H_y\approx 0$. Then Eq.~(\ref{eq17})
written for the in-plane magnetization leads to
\begin{eqnarray}
&&
H_x(z_2)=\tilde{M}_{00}^{(1)}
\int_{-L_x^{(1)}/2}^{L_x^{(1)}/2}\!\!\!dx_1
\int_{-L_y^{(1)}/2}^{L_y^{(1)}/2}\!\!\!dy_1
\int_{-d_1}^{0}\!\!\!dz_1
\left\{
\frac{3x_1^2}{\left[x_1^2+y_1^2+(z_2-z_1)^2\right]^{5/2}}
\right.
\nonumber \\
&&
~~~
\left.-
\frac{1\vphantom{x_1^2}}{\left[x_1^2+y_1^2+(z_2-z_1)^2\right]^{3/2}}
\right\}.
\label{B12}
\end{eqnarray}
\noindent
Using the identity
\begin{equation}
\frac{3x_1^2}{\left[x_1^2+y_1^2+(z_2-z_1)^2\right]^{5/2}}=
-x_1\frac{\partial}{\partial x_1}
\frac{1}{\left[x_1^2+y_1^2+(z_2-z_1)^2\right]^{3/2}},
\label{B13}
\end{equation}
\noindent
we calculate the integral with respect to $x_1$ and obtain
\begin{equation}
H_x(z_2)=8\tilde{M}_{00}^{(1)}
L_x^{(1)}
\int_{-L_y^{(1)}/2}^{L_y^{(1)}/2}\!\!\!dy_1
\int_{-d_1}^{0}\!\!\!dz_1
\frac{1}{\left[(L_x^{(1)})^2+4y_1^2+4(z_2-z_1)^2\right]^{3/2}}.
\label{B14}
\end{equation}
\noindent
Both integrations in Eq.~(\ref{B14}) can be easily performed with
the result
\begin{equation}
H_x(z_2)=4\tilde{M}_{00}^{(1)}
\left[\arctan\frac{2(z_2+
d_1)}{\sqrt{(L_x^{(1)})^2+(L_y^{(1)})^2+4(z_2+d_1)^2}}-
\arctan\frac{2z_2}{\sqrt{(L_x^{(1)})^2+(L_y^{(1)})^2
+4z_2^2}}\right].
\label{B15}
\end{equation}
\noindent
Substituting Eq.~(\ref{B15}) in the following expression for
the magnetic energy per unit area due to the spontaneous
magnetization:
\begin{equation}
E_{ms}(a)=-\tilde{M}_{x;00}^{(2)}
\int_{a}^{a+d_2}\!\!\!dz_2H_x(z_2),
\label{B16}
\end{equation}
\noindent
one arrives at Eq.~(\ref{eq39}).


\begin{figure}[b]
\vspace*{-6cm}
\centerline{\hspace*{3cm}
\includegraphics{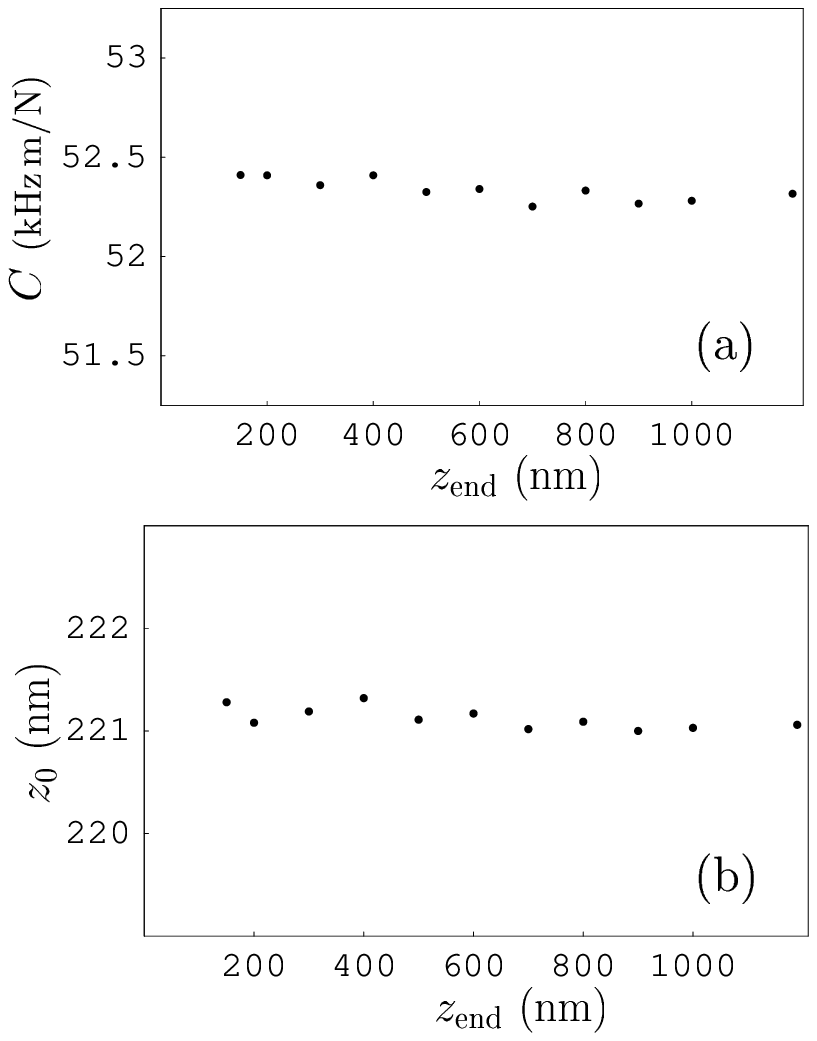}
}
\vspace*{-11.5cm}
\caption{\label{fg1}
(a) The coefficient $C$ in Eq.~(\ref{eq8}) and (b) the
closest sphere-plate separation $z_0$ as functions of the end
point of the fit.
}
\end{figure}
\begin{figure}[b]
\vspace*{-15cm}
\centerline{\hspace*{3cm}
\includegraphics{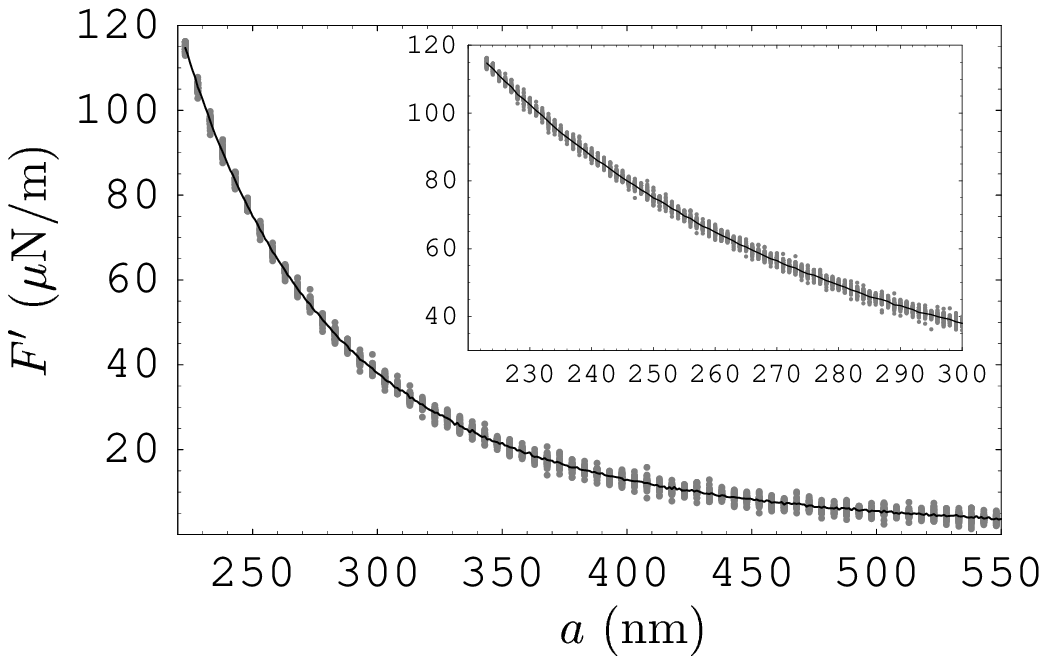}
}
\vspace*{-7cm}
\caption{\label{fg2}
All 33 data points for the gradient of the Casimir force
between Ni surfaces are shown as dots with a step of 5\,nm
starting from a separation of 223\,nm. The mean values of the
measured gradients are presented as the solid line.
In the inset the same information is given with a step of
1\,nm over a more narrow region.
}
\end{figure}
\begin{figure}[b]
\vspace*{-15cm}
\centerline{\hspace*{3cm}
\includegraphics{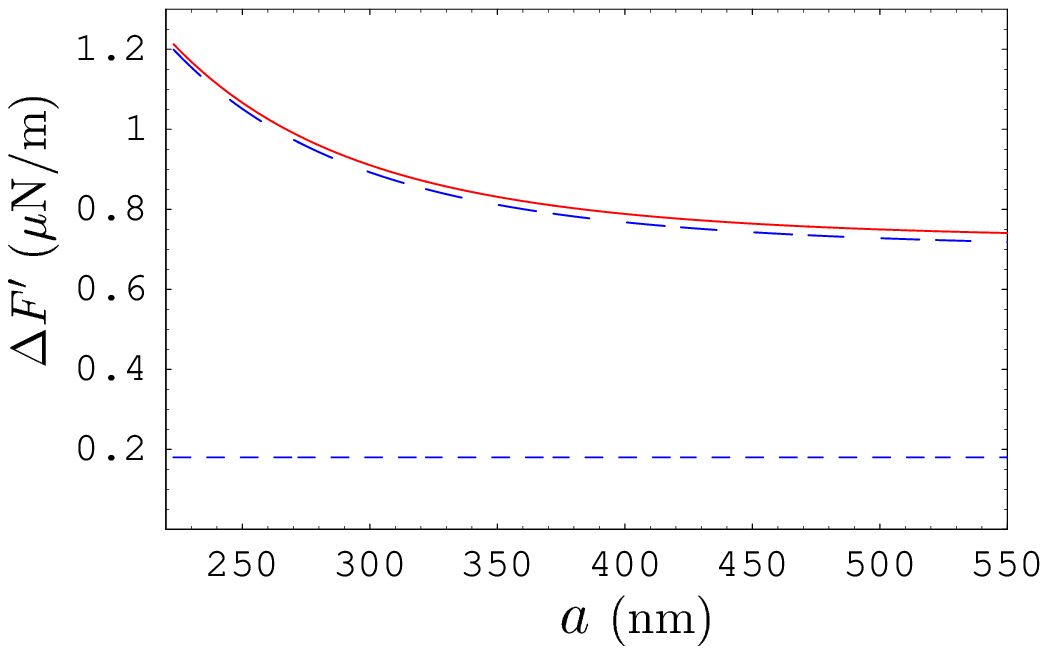}
}
\vspace*{-7cm}
\caption{\label{fg3}(Color online)
The random, $\Delta^{\! r}F^{\prime}$, systematic,
$\Delta^{\! s}F^{\prime}$, and total,  $\Delta^{\! t}F^{\prime}$,
errors in the measured gradient of the Casimir force determined
at a 67\% confidence level are shown as functions of separation
by the short-dashed, long-dashed, and solid lines, respectively.
}
\end{figure}
\begin{figure}[b]
\vspace*{-15cm}
\centerline{\hspace*{3cm}
\includegraphics{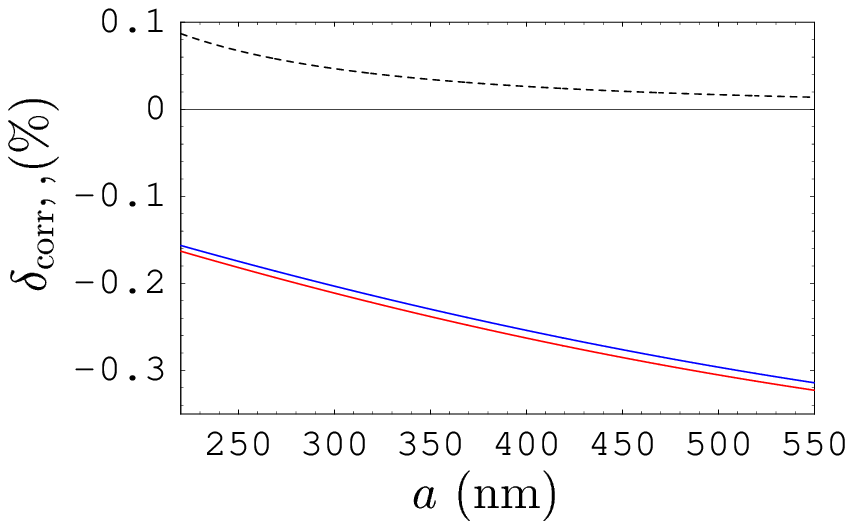}
}
\vspace*{-7cm}
\caption{\label{fg4}(Color online)
Corrections to the gradient of the Casimir force due to
deviations from the PFA (upper and lower solid lines
computed within the Drude and plasma model approaches,
respectively) and due to the surface roughness (dashed line)
as functions of separation.
}
\end{figure}
\begin{figure}[b]
\vspace*{-15cm}
\centerline{\hspace*{3cm}
\includegraphics{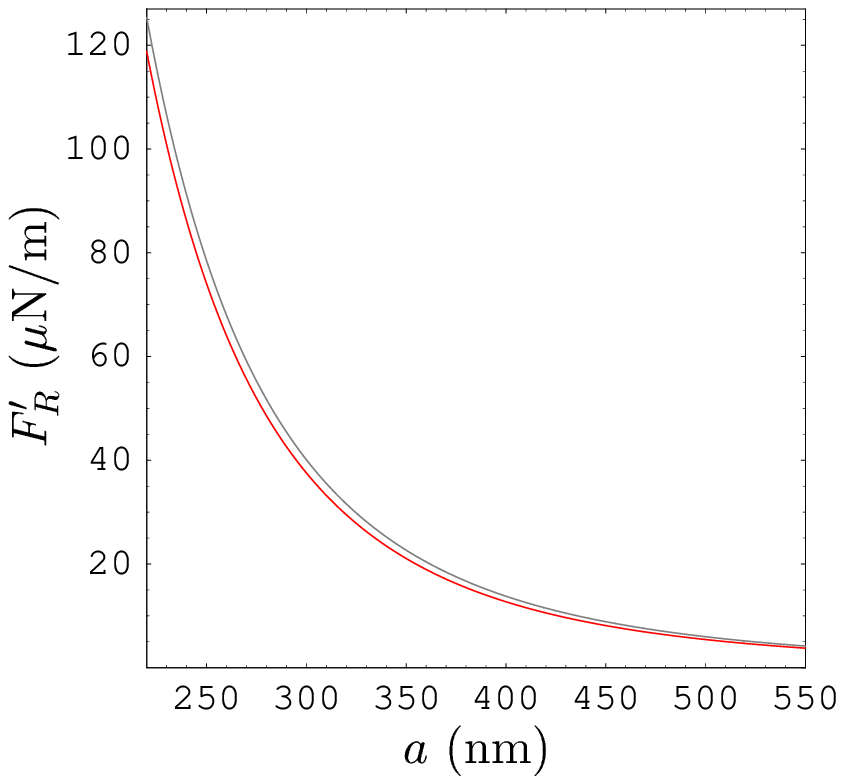}
}
\vspace*{-7cm}
\caption{\label{fg5}(Color online)
Theoretical predictions for the gradient of the Casimir
force between Ni surfaces computed using the Drude and
plasma model approaches (upper and lower lines,
respectively) including corrections to the PFA and due to
surface roughness as functions of separation.
}
\end{figure}
\begin{figure}[b]
\vspace*{-6cm}
\centerline{\hspace*{3cm}
\includegraphics{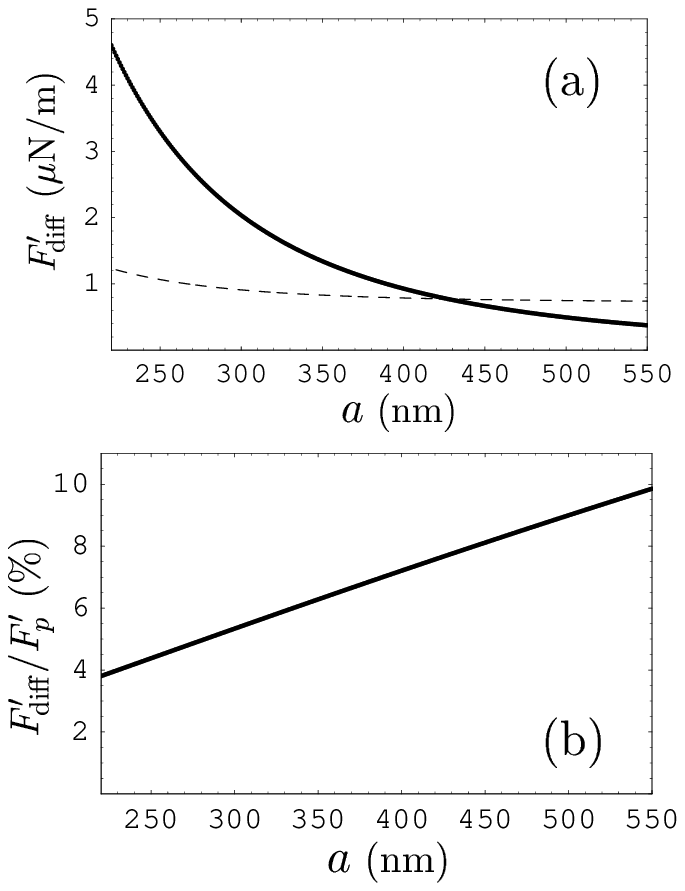}
}
\vspace*{-11.5cm}
\caption{\label{fg6}
(a) The difference of gradients of the Casimir
force between Ni surfaces predicted within the Drude and
plasma model approaches as a function of separation
is shown by the solid line (the dashed
line indicates the total experimental error determined
at a 67\%  confidence level).
(b) The relative difference of force gradients
predicted within the Drude and
plasma model approaches as a function of separation.
}
\end{figure}
\begin{figure}[b]
\vspace*{-15cm}
\centerline{\hspace*{3cm}
\includegraphics{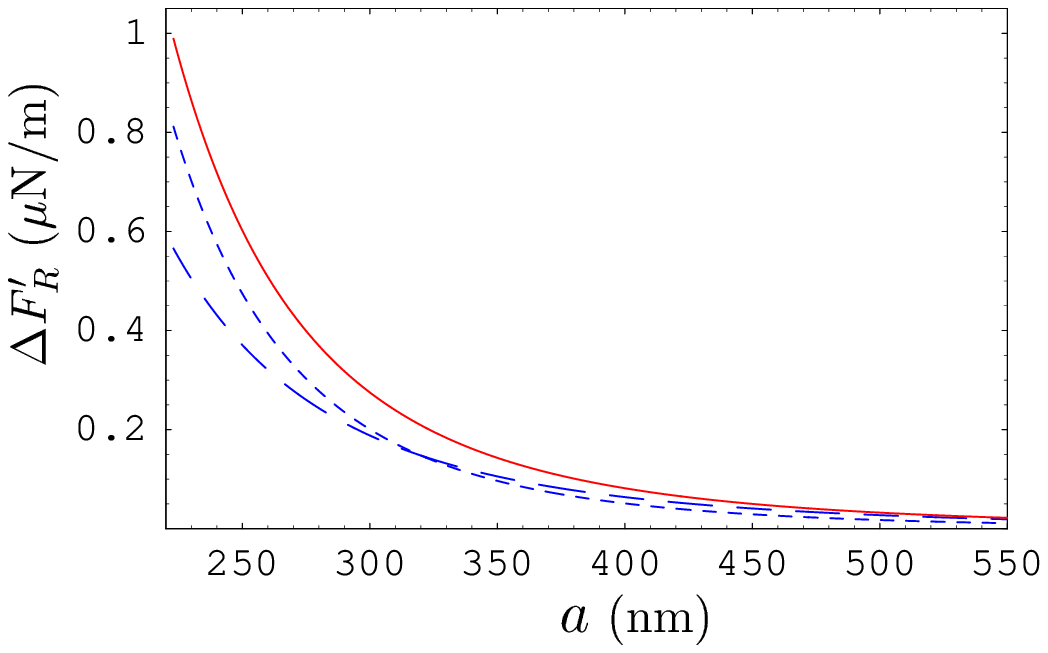}
}
\vspace*{-7cm}
\caption{\label{fg7}(Color online)
The errors in the theoretical gradient of the Casimir force
$F_{R}^{\prime}(a_i)$ due to inaccuracy of the optical
data of Ni, $\Delta^{\!\rm opt}F_{R}^{\prime}$,
due to  the errors in measured separations $a_i$,
$\Delta^{\!\rm sep}F_{R}^{\prime}$,
and the total theoretical error,
$\Delta^{\! t}F_{R}^{\prime}$, are
shown as functions of separation by the long-dashed,
short-dashed, and solid lines, respectively.
}
\end{figure}
\begin{figure}[b]
\vspace*{-6cm}
\centerline{\hspace*{3cm}
\includegraphics{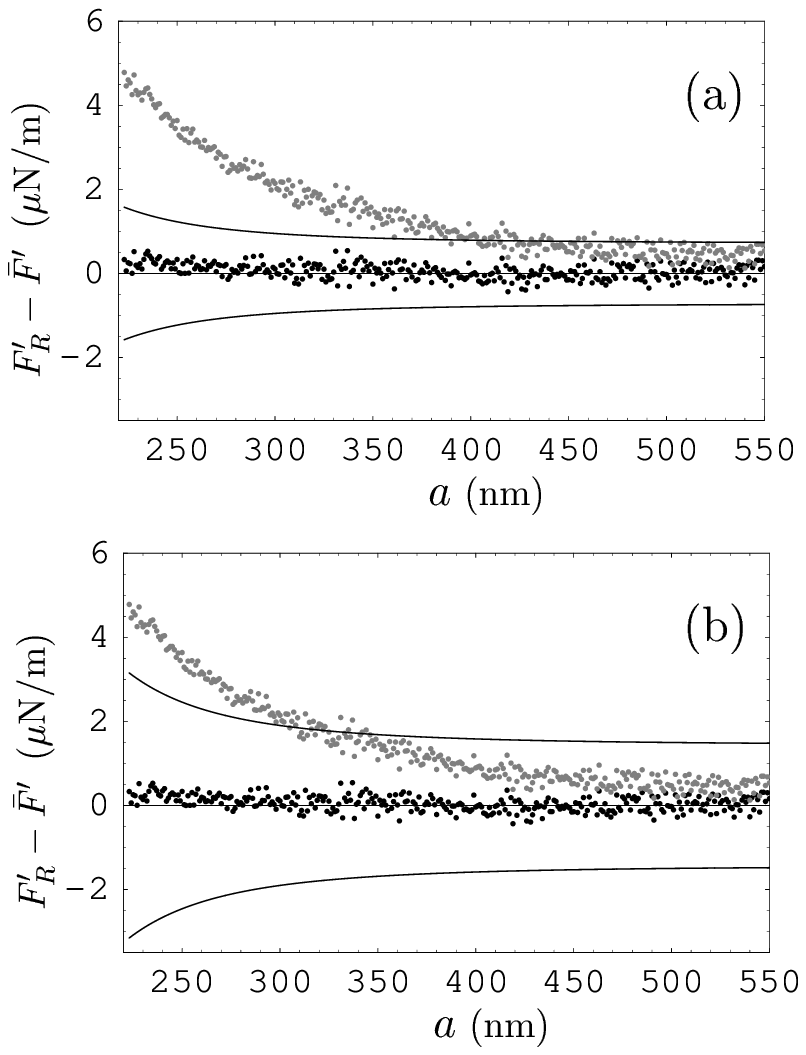}
}
\vspace*{-11.5cm}
\caption{\label{fg8}
Differences between the theoretical and mean experimental
gradients of the Casimir force found at the experimental
separations using the plasma  and the Drude model approaches
are shown by the black and gray dots, respectively.
The solid lines indicate the borders of the (a) 67\% and
(b) 95\% confidence intervals.
}
\end{figure}
\begin{figure}[b]
\vspace*{-13cm}
\centerline{\hspace*{3cm}
\includegraphics{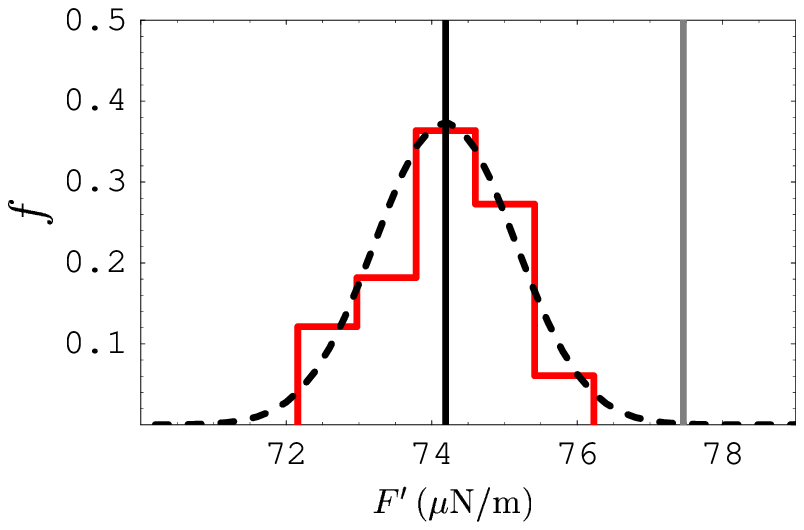}
}
\vspace*{-10cm}
\caption{\label{fg9}(Color online)
Histogram for the measured gradient of the Casimir force at the
separation $a=250\,$nm (see text for details). The theoretical
predictions of the plasma and Drude model approaches are shown
by the black and gray vertical lines, respectively.
}
\end{figure}
\begin{figure}[b]
\vspace*{-15cm}
\centerline{\hspace*{3cm}
\includegraphics{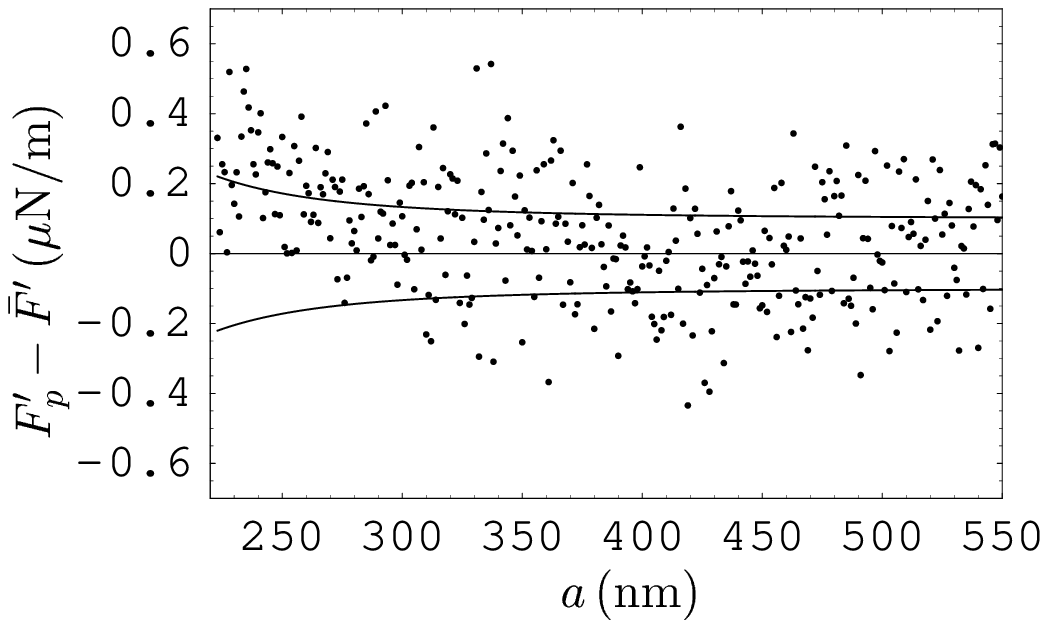}
}
\vspace*{-7cm}
\caption{\label{fg10}
Differences between the theoretical and mean experimental
gradients of the Casimir force found at the experimental
separations using the plasma model approach
are shown by the black dots.
The solid lines indicate the borders of the 10\%
confidence intervals.
}
\end{figure}
\begin{figure}[b]
\vspace*{-6cm}
\centerline{\hspace*{3cm}
\includegraphics{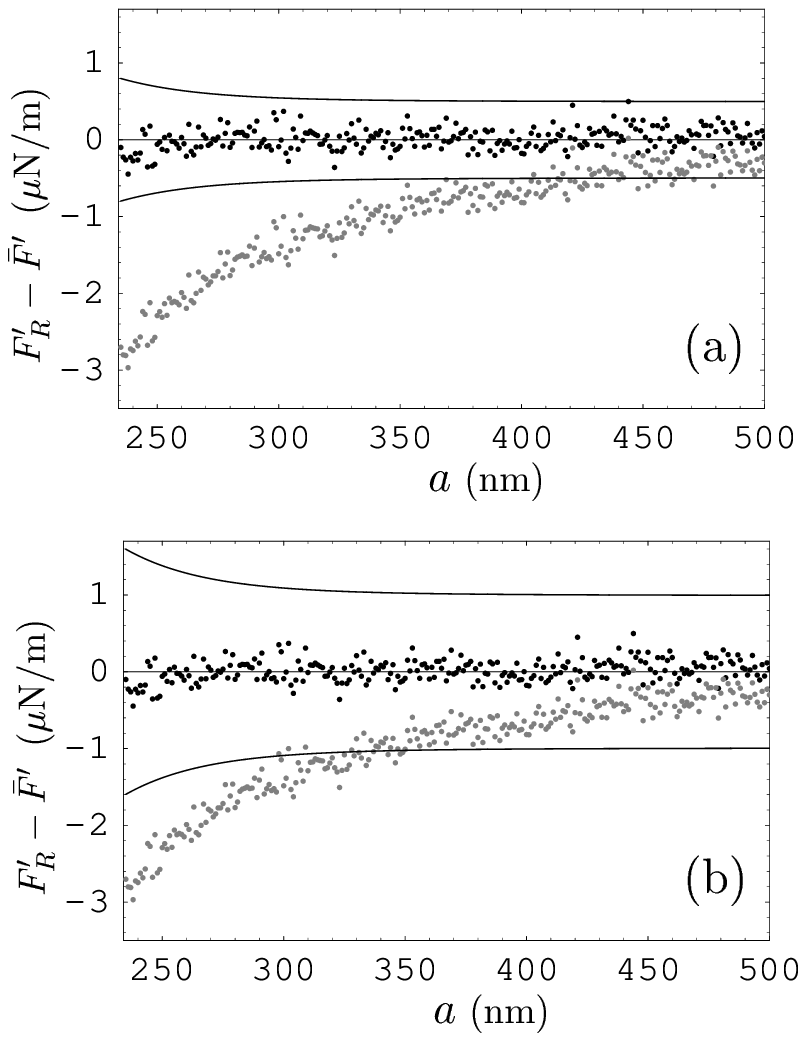}
}
\vspace*{-11.5cm}
\caption{\label{fg11}
Differences between the theoretical and mean experimental
gradients of the Casimir force found at the experimental
separations between a plate and a sphere both coated with
Au using the plasma  and the Drude model approaches
are shown by the black and gray dots, respectively.
The solid lines indicate the borders of the (a) 67\% and
(b) 95\% confidence intervals.
}
\end{figure}
\begin{figure}[b]
\vspace*{-6cm}
\centerline{\hspace*{3cm}
\includegraphics{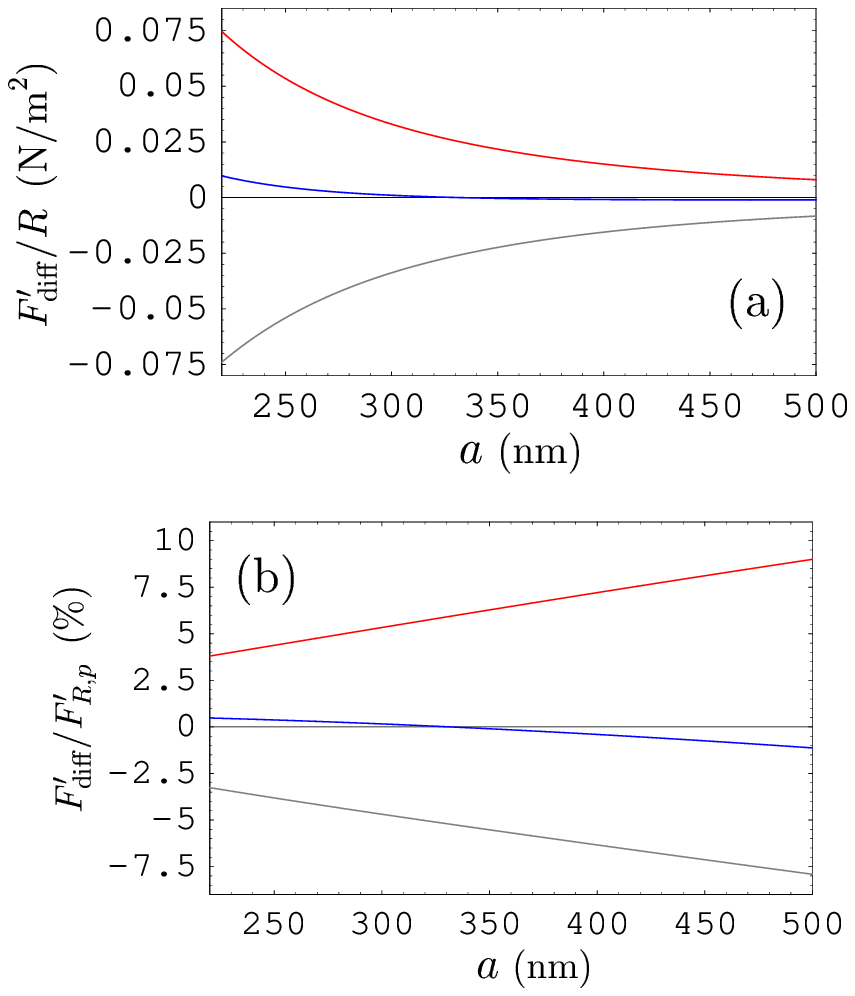}
}
\vspace*{-11.5cm}
\caption{\label{fg12}(Color online)
Difference of gradients of the Casimir
force calculated using  the Drude and
plasma model approaches and normalized (a) for
the respective
sphere radii and (b) for the gradients of the Casimir
force calculated using  the plasma model approach
are shown by the solid lines from top to bottom for
experiments with Ni-Ni, Ni-Au and Au-Au test bodies,
respectively.
}
\end{figure}
\begin{figure}[b]
\vspace*{-4cm}
\centerline{\hspace*{3cm}
\includegraphics{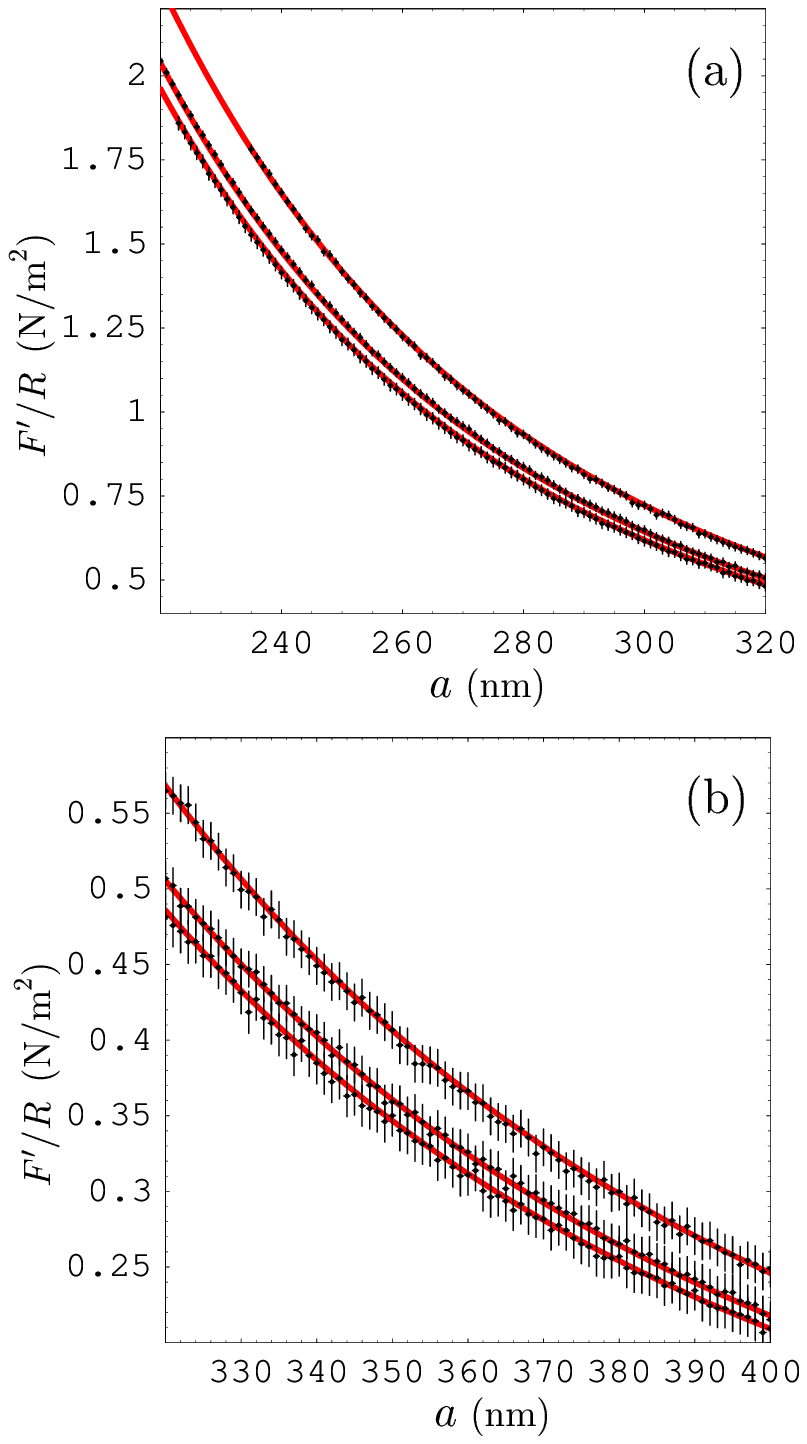}
}
\vspace*{-9.5cm}
\caption{\label{fg13}(Color online)
The measurement data for the mean gradients of the Casimir force
normalized by sphere radii with the total experimental errors
indicated as crosses and theoretical bands computed using the
plasma model approach are shown from top to bottom for
the experiments with Au-Au, Au-Ni and Ni-Ni surfaces over the
separation region (a) from 220 to 320\,nm and (b) from 320
to 400\,nm.
}
\end{figure}
\begin{figure}[b]
\vspace*{-4cm}
\centerline{\hspace*{3cm}
\includegraphics{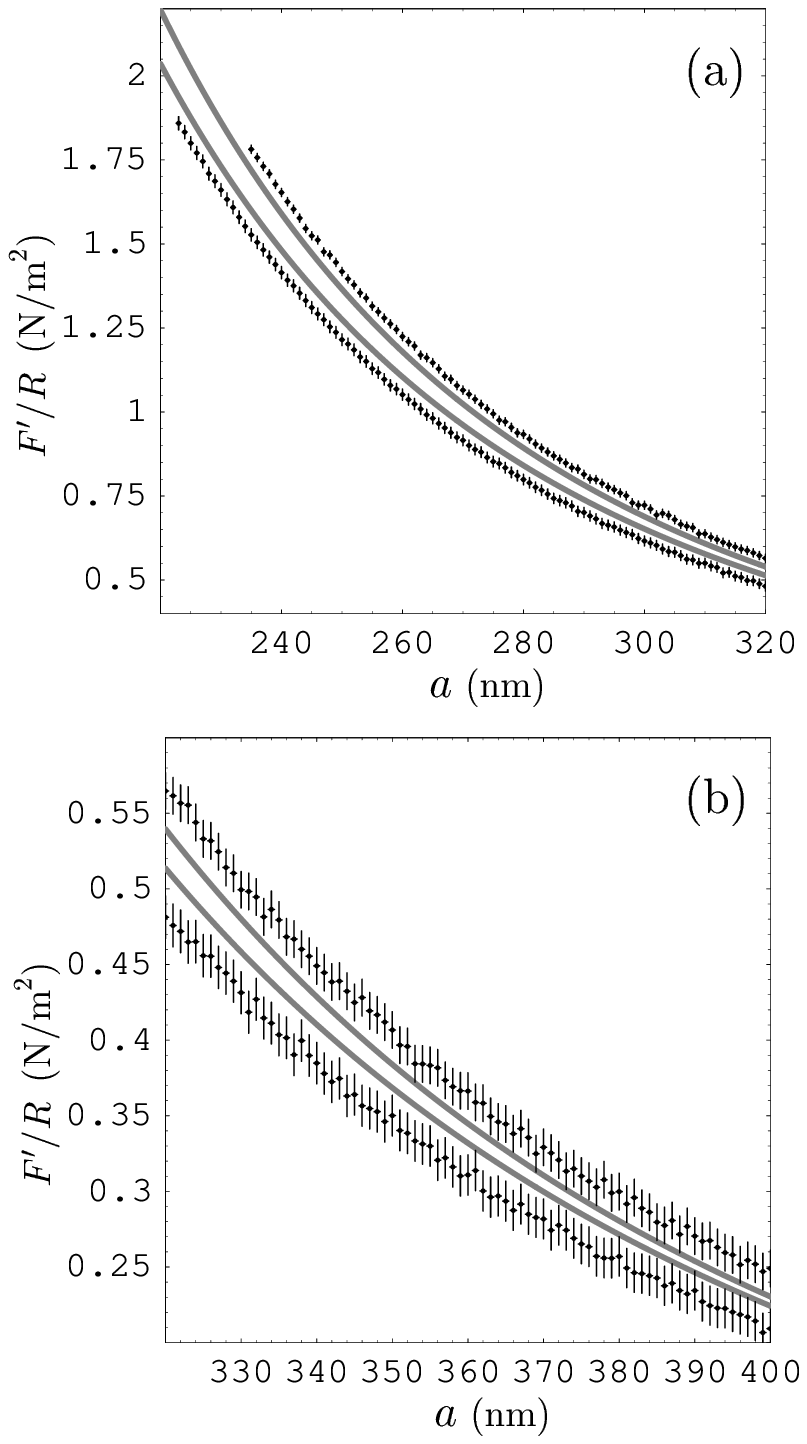}
}
\vspace*{-9.5cm}
\caption{\label{fg14}
The measurement data for mean gradients of the Casimir force
normalized by sphere radii with total experimental errors
indicated as crosses and the theoretical bands computed using the
Drude model approach are shown from top to bottom for
the experiments with Au-Au and Ni-Ni surfaces over the
separation region (a) from 220 to 320\,nm and (b) from 320
to 400\,nm.
}
\end{figure}

\end{document}